\documentclass[11pt,a4paper]{article}

\usepackage{geometry}
\usepackage{helvet} 

\usepackage{placeins}
\usepackage[utf8x]{inputenc} 
\usepackage{multicol}
\usepackage{csquotes}
\usepackage{indentfirst}

\usepackage{graphicx}
\usepackage{times}
\usepackage[bottom]{footmisc}	
\usepackage{amsmath, amssymb, amsthm, graphicx, color, bm, soul, array}
\usepackage{amsmath}
\usepackage{authblk}
\usepackage{xcolor}			
\usepackage{caption}
\usepackage{tikz}

\usepackage{multirow}

\usepackage{qcircuit}
\usepackage{physics}
\usepackage{braket} 

\usepackage{musicography}  

\begin{document}
\begingroup
\let\center\flushleft
\title{\textbf{\text{\LARGE{QuiKo: A Quantum Beat Generation Application}}}}
\author[]{\color{gray}\large{\textbf{Scott Oshiro}}}
\affil[]{\color{gray}\large{\textbf{CCRMA, Stanford University, USA}}}
\date{}
\maketitle
\let\endcenter\endflushleft
\endgroup

\captionsetup[figure]{font=footnotesize, labelfont=bf}
\footnotetext[1]{Pre-Publication Chapter (Unedited), to appear in the book "Quantum Computer Music", E.R. Miranda (Ed.)}%


\section*{Introduction}
\medskip
Artificial intelligent (AI) music generation systems have been exciting developments in machine and deep learning, but are limited to the data set(s) that they are fed. As a result, these systems lack a sense of organicness, or intuition, in their responses to external musical events. It has been speculated that quantum computing can be leveraged to go beyond just the imitation of the provided data set to the system. But concrete methods and results have not yet been presented to support this concept. However, Quantum Machine learning (QML) algorithms [1] can be dissected and adapted to begin developing algorithms that could possibly give these AI music generation systems the organic touch that they need.\hfill

\medskip
In this chapter a quantum music generation application called QuiKo will be discussed. It combines existing quantum algorithms with data encoding methods from QML [1] to build drum and audio sample patterns from a database of audio tracks. QuiKo leverages the physical properties and characteristics of quantum computers to generate what can be referred to as Soft Rules proposed by Kirke, A.[2]. These rules take advantage of noise produced by the quantum devices to develop flexible rules and grammars for quantum music generation. These properties include qubit decoherence and phase kickback due controlled quantum gates within the quantum circuit.\hfill

\medskip
QuiKo builds upon the concept of soft rules in quantum music generation and takes it a step further. It attempts to mimic and react to an external musical inputs, similar to the way that human musicians play and compose with one another. Audio signals (ideally rhythmic in nature) are used as inputs into the system. Feature extraction is then performed on the signal to identify it’s harmonic and percussive elements. This information is then encoded onto QuiKo's quantum algorithm’s quantum circuit. Then measurements of the quantum circuit are taken providing results in the form of probability distributions for external music applications to use to build the new drum patterns.\hfill

\medskip
In Section I, the system overview of the QuiKo application will be covered while in section II walks through the several quantum algorithms that act as building blocks for the application. Section III \& IV will then outline in detail the inner workings of QuiKo along with the two different encoding methods. Section V \& VI will then present results and analysis of the performance of the QuiKo circuit. They will also discuss initial experiments in building out the whole application in one quantum circuit. Finally, section VII will discuss future work.\hfill

\section*{System Overview} 
\medskip
Quiko, developed using IBM's quantum framework, Qiskit [3], and has three main components (1) Preprocessing (2) Quantum Circuit (3) Beat Construction. Elements within these components are flexible and can be modified by the user, but we will stick to specific methods presented in this chapter.\hfill

\begin{figure}[htbp]
\begin{center}\vspace{0.2cm}
\includegraphics[width=0.9\linewidth]{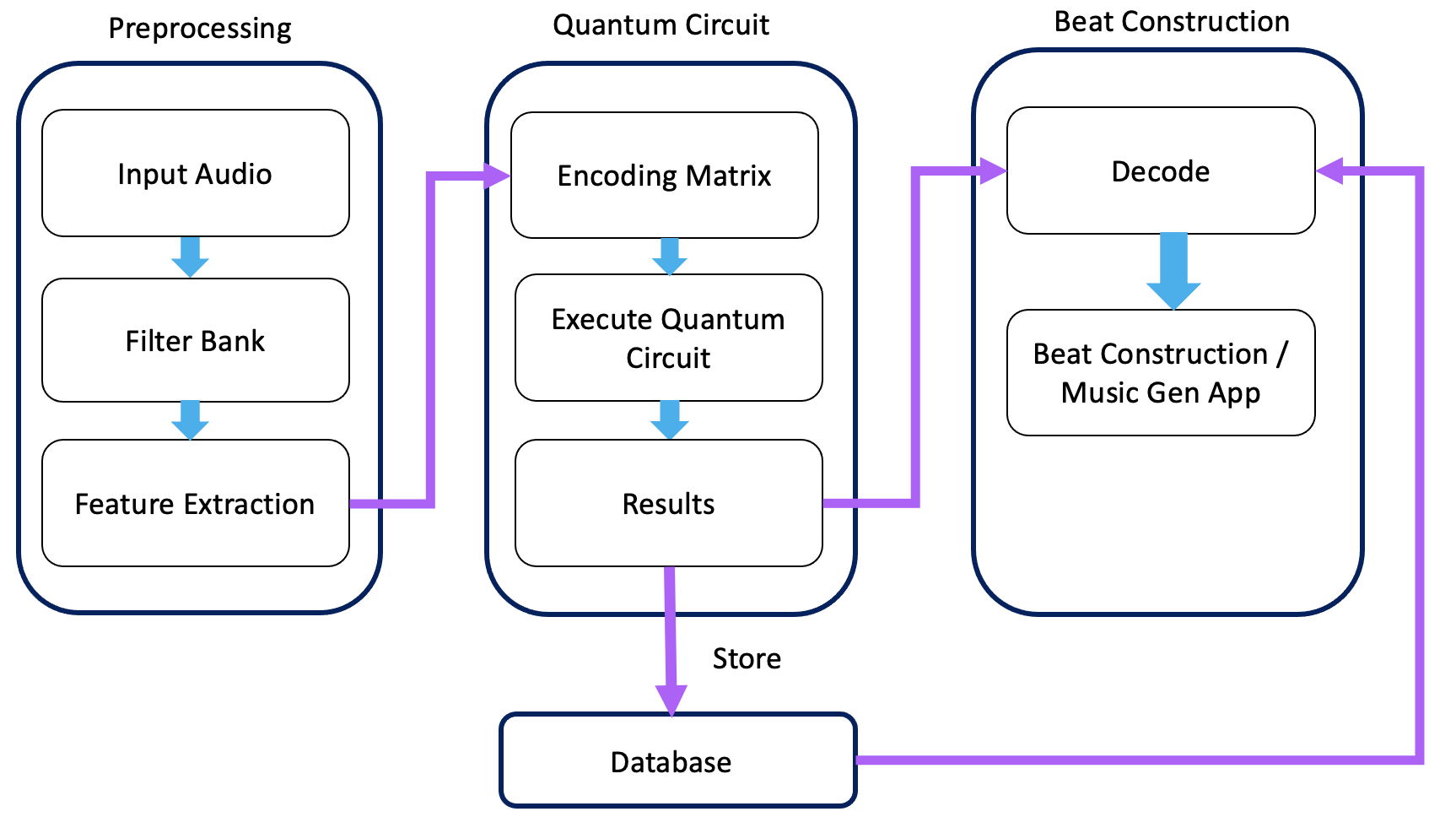}
\caption{QuiKo Architecture}
\end{center}
\end{figure}
\FloatBarrier

\medskip
First, the pre-processing component takes in an audio file, containing a piece of music, and extracts specific features from it. This provides a guide for how the system should aim to generated the new beat. It acts as a sort of influence or template for the system to use. To do this, the input audio file is fed into a filter bank producing filtered versions of the original audio based on a specific sub-band mappings. For simplicity, we will be using three filters. One for low frequency content (low pass), one for mid frequency (bandpass) content and one for high frequency (high pass) content, giving a total of three bands. The purpose of this step will become more clear in later sections. The system then performs feature extraction for collecting specific musical data for each measure and subdivision in the audio file.\hfill

\medskip
We then move to the second component of the system, which is the encoder. Here the features extracted in the pre-processing module are encoded onto the quantum circuit using controlled Unitary quantum gates (U gates), which we will discuss future sections. First, the encoder organizes the data into an encoding matrix in order to easily access and encoded the extracted features onto their corresponding Unitary gates based on a specific subdivisions. It then building out the core quantum circuit to be used in the generation of a new beat. Here we will discuss two methods, static encoding and phase kick back sequencing encoding (PKBSE). The circuit is initialized and measured for 1024 times (shots). The results are obtained and recorded for each shot.\hfill

\medskip
The third and final component includes the decoder and beat constructor. After we have collected the results from running our quantum circuit, this component parses out the results for each subdivision and compares the states of the input audio to the states associated with the audio files in the database. From there, the system can determine which audio files (samples) in the database are more or less correlated with the quantum state of the input audio file. This information is then fed into a music generation application, developed in another computer music framework or DAW such as WebAudio API, MAX MSP, Abelton, Logic Pro, etc, to build out the final beat. Currently, separate circuits are needed to be run for each audio file in the database to obtain their resulting probability distributions. Thus, the comparison between the audio files in the database and the input audio is performed classically. However, in the future work section, designs and findings are presented from initial experiments in combining the quantum circuit and comparison process into one quantum circuit.\hfill

\section*{Algorithm Building Blocks}
\medskip
Before we dive into the specifics of the design for this application, we first need to discuss the underlying quantum algorithms and properties that are being utilized. These primarily include the Quantum Fourier Transform (QFT) [4] and Quantum Phase Estimation (QPE) [4]. These will be used to handle the rhythmic elements of the output beat, while the timbre and spectral elements will be handled using methods similar to amplitude and phase encoding used in quantum machine learning (QML) [1].\hfill

\subsection*{Quantum Fourier Transform (QFT)}

\medskip
The Quantum Fourier Transform (QFT) lies at the heart of many different quantum algorithms such as phase estimation along with Shor's factoring and period finding algorithms [4]. Essentially, the QFT transforms our states from the computational basis to Fourier Basis. We can gain some intuition of this by studying the bloch's sphere in figure 2. If we assume the qubit is initialized in the ground state $\ket{0}$ and we then apply a hadamard gate to the qubit we transforms its state from $\ket{0}$ to a state of equal superposition between 0 and 1. In other words we rotate our statevector from the north pole of the Bloch's sphere to the equator. This, changes our basis states from $\ket{0}$ and $\ket{1}$ to $\ket{+}$ and $\ket{-}$ in the Fourier basis.\hfill

\begin{figure}[htbp]
\begin{center}\vspace{0.2cm}
\includegraphics[scale=0.5]{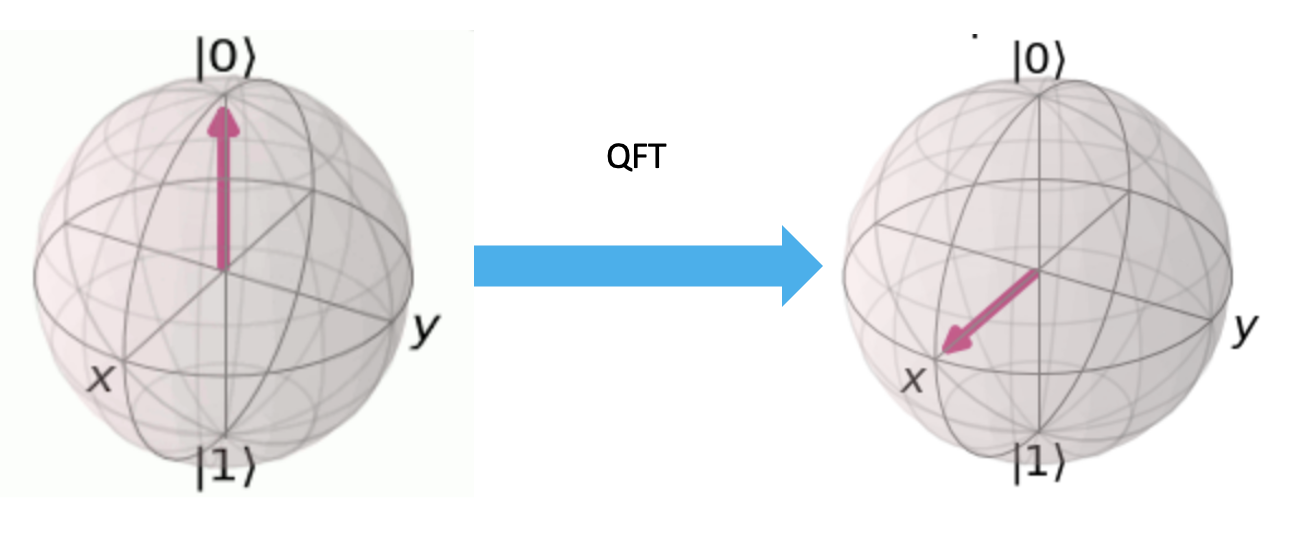}
\caption{Percussive and Harmonic parts of the Audio Signal}
\end{center}
\end{figure}
\FloatBarrier

Mathematically, we can express this transform for a single qubit as follows:\hfill

\begin{equation}
\tilde{\ket{X}} = QFT\ket{X} = \frac{1}{\sqrt{N}}\sum_{k=0}^{N-1} \omega^{j_k}_N \ket{k}
\end{equation}

where $\omega^{j_k}_N = e^\frac{2\pi ixy}{N}$. If we were to apply this to the single qubit case we would get:\hfill

\begin{align*}
\tilde{\ket{0}} = QFT\ket{0} = \frac{1}{\sqrt{2}}\sum_{k=0}^{N-1} \omega^{j_k}_N \ket{k} = \frac{1}{\sqrt{2}}(e^\frac{2\pi i(0)(0)}{2}\ket{0} + e^\frac{2\pi i(0)(1)}{2}\ket{1}) 
\end{align*}

\begin{align*}
= \frac{1}{\sqrt{2}}(\ket{0} + \ket{1})
\end{align*}

\begin{align*}
\tilde{\ket{1}} = QFT\ket{1} = \frac{1}{\sqrt{2}}\sum_{k=0}^{N-1} \omega^{j_k}_N \ket{k} = \frac{1}{\sqrt{2}}(e^\frac{2\pi i(1)(0)}{2}\ket{0} + e^\frac{2\pi i(1)(1)}{2}\ket{1}) = \frac{1}{\sqrt{2}}(\ket{0} + e^{\pi i}\ket{1}
\end{align*}

\begin{align*}
= \frac{1}{\sqrt{2}}(\ket{0} - \ket{1})
\end{align*}

The implementation of the QFT becomes more complex as we scale up to more qubits due to the fact we have more states to deal with. After we put all qubits in equal superposition we can then encode different values within their phases. We can encode information by rotating the state of each qubit by a different amount around the equator of the Bloch's sphere. The rotation of each qubit depends on the angle of rotation of the other qubits. For example, to encode some state $\tilde{\ket{x}}$ on 3 qubits we will need to rotate the least significant bit (LSB) by $\frac{x}{2^n}$, which in this case would be $\frac{x}{2^3}$ = $\frac{x}{8}$ full turns. The next qubit would then have to rotate twice as much, and so on an so forth depending on the number of qubits. As a result, the circuit for the QFT is going to implement a series of controlled Z gates in order to appropriately entangle the qubits being transformed to the fourier basis.\hfill

\medskip
This process may seem a little intimidating, but mathematically, we can break it down into individual qubit parts to make it easier for us to understand. If we have n qubits we have $N = 2^n$ states. Let's say for example we have 3 qubits, $n = 3$, and as a result have $N = 2^3 = 8$ states. Our states in the computational basis is going to look like:

\begin{equation}
\ket{y_1 y_2 ... y_n} = 2^{n-1} y_{1} + 2^{n-2} y_{2} + . . . + 2^{0} y_{n} = \sum_{k=1}^{n} y_k 2^{n-k}
\end{equation}

Which is just how we would represent a specific state in binary such as $\tilde{\ket{7}} = \tilde{\ket{111}}$. Each $y_n$ represents a single bit in the binary string. If we plug this into the QFT equation we defined earlier we get:

\begin{equation}
\tilde{\ket{x}} = \frac{1}{\sqrt{N}} \sum_{y=0}^{N-1} e^{i2\pi x\sum_{k=1}^{n} y_k 2^{n-k}} \ket{y_1 y_2 ... y_n} = \frac{1}{\sqrt{N}} \sum_{y=0}^{N-1} \prod_{y=0}^{n} e^{\frac{2\pi ixy_k}{2^k}}\ket{y_1 y_2 ... y_n}
\end{equation}

\begin{equation}
\tilde{\ket{x}} = \frac{1}{\sqrt{N}}(\ket{0} + e^{\frac{2\pi ix}{2^1}}\ket{1}) \otimes (\ket{0} + e^{\frac{2\pi ix}{2^2}}\ket{1}) \otimes (\ket{0} + e^{\frac{2\pi ix}{2^3}}\ket{1}) \otimes ... \otimes (\ket{0} + e^{\frac{2\pi ix}{2^n}}\ket{1})
\end{equation}

We can know expand out the equation (3) so that we have the tensor products of qubit rotating at the a specific angle that we have specified in relation to the other qubits as seen in equation (4). We can think of the first parenthesis as the LSB while the elements in last parenthesis represents the state of the qubit in the MSB position. Also we can also observe that the rotations are applying a global phase on each of the individual qubits as the $e^{\frac{2\pi ix}{2^n}}$ elements.\hfill

Looking at equation (4), we can build out the circuit for the $QFT$ on a multi-qubit system as follows:\hfill

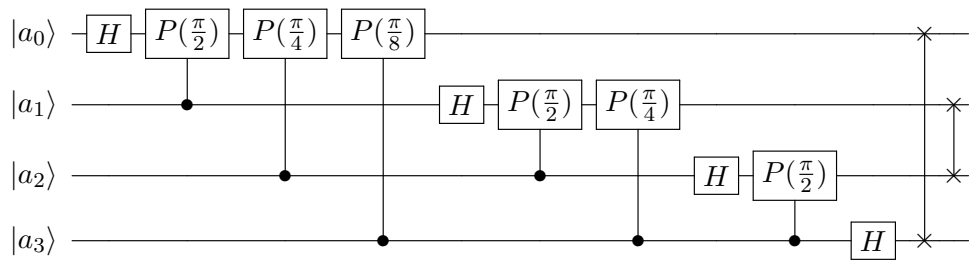
\begin{figure}[htbp]
\begin{align*}
\Qcircuit @C=0.5em @R=0.75em {
&\lstick{\ket{a_0}}  & \gate{H} & \gate{P(\frac{\pi}{2})} & \gate{P(\frac{\pi}{4})} & \gate{P(\frac{\pi}{8})} & \qw & \qw & \qw & \qw & \qw & \qw & \qw & \qswap & \qw & \qw & \qw\\
&\lstick{\ket{a_1}}  & \qw  & \ctrl{-1} & \qw & \qw & \gate{H} & \gate{P(\frac{\pi}{2})} & \gate{P(\frac{\pi}{4})} & \qw & \qw & \qw & \qw & \qw \qwx & \qw & \qswap & \qw\\
&\lstick{\ket{a_2}}  & \qw  & \qw & \ctrl{-2} & \qw & \qw & \ctrl{-1} & \qw & \gate{H} & \gate{P(\frac{\pi}{2})} & \qw & \qw & \qw \qwx & \qw & \qswap \qwx & \qw\\
&\lstick{\ket{a_3}}  & \qw  & \qw & \qw & \ctrl{-3} & \qw & \qw & \ctrl{-2} & \qw & \ctrl{-1} & \gate{H} & \qw & \qswap \qwx & \qw & \qw & \qw\\
}
\end{align*}
\caption{Quantum Circuit for Quantum Fourier Transform ($QFT$)}
\end{figure}

\medskip
However, if we were to measure the circuit as is, after the forward $QFT$ we would get results identical to the equal superposition case of all qubits in the register. This is because all qubits are in a state of equal superposition of $\ket{0}$ and $\ket{1}$. If we want to make this useful we would have to encode a specific state on to the phase of the qubit register in the phase domain (Fourier basis) and then apply what is called in the inverse $QFT$ or $QFT^\dagger$. This transforms the Fourier basis back into the computational basis. This circuit can be implement simply by reversing the $QFT$ operation. The quantum circuit is illustrated in figure 4.\hfill

\begin{figure}[!htb]
\begin{center}
\includegraphics[scale=0.75]{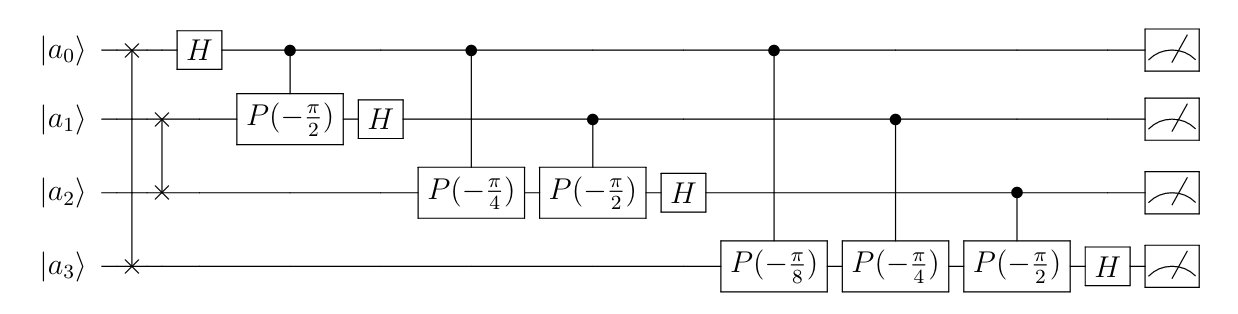}
\caption{Quantum Circuit for Inverse Quantum Fourier Transform $QFT^{\dagger}$}
\end{center}
\end{figure}

\medskip
The $QFT^\dagger$ is useful in quantum algorithms that need to perform operations in the fourier basis such as addition and multiplication as presented in [5]. More commonly, the practical use of the $QFT^\dagger$ is used within the quantum phase estimation algorithm.\hfill

\subsection*{Quantum Phase Estimation PE}
Quantum Phase Estimation demonstrates the practical use cases for the QFT and $QFT^\dagger$. This algorithm estimates the amount of phase that an unitary gate applies to a qubit. Let's consider the following quantum circuit below as an example. This example is outlined the qiskit text book [4]

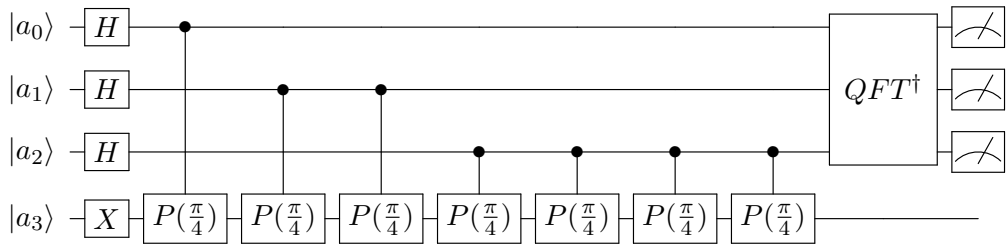
\begin{figure}[htbp]
\begin{align*}
\Qcircuit @C=0.50em @R=0.75em {
&\lstick{\ket{a_0}} & \gate{H} & \ctrl{3} & \qw & \qw & \qw & \qw & \qw & \qw & \multigate{2}{QFT^\dag} & \meter\\
&\lstick{\ket{a_1}} & \gate{H} & \qw & \ctrl{2} & \ctrl{2} & \qw & \qw & \qw & \qw & \ghost{QFT^\dag} & \meter\\
&\lstick{\ket{a_2}} & \gate{H} & \qw & \qw & \qw & \ctrl{1} & \ctrl{1} & \ctrl{1} & \ctrl{1} & \ghost{QFT^\dag} & \meter\\
&\lstick{\ket{a_3}} & \gate{X} & \gate{P(\frac{\pi}{4})} & \gate{P(\frac{\pi}{4})} & \gate{P(\frac{\pi}{4})}& \gate{P(\frac{\pi}{4})}& \gate{P(\frac{\pi}{4})}& \gate{P(\frac{\pi}{4})}& \gate{P(\frac{\pi}{4})} & \qw & \qw\\
}
\end{align*}
\caption{Quantum Phase Estimation Quantum Circuit}
\end{figure}

\medskip
This controlled unitary gate, U-gate, applies a specific amplitude and phase to the target qubit. However, the phase applied by the U-gate to the target qubit also get kicked back and applied to the control qubit. This effect is called phase kickback. In order to estimate the phase of the unitary we need to apply full turns on the MSB. We will use cascading controlled phase gates (P gates) to create these rotations. This circuit is illustrated in figure 5. We use an auxiliary qubit $a_3$ to apply the P gates while the phases of those gates are kicked back to their control qubit. The circuit above shows that we rotate LSB by $\pi/4$ and then $a_1$ by $\pi/2$ and $a_2$ by $\pi$ due to the phase kickback. This is similar to what we have seen from the previous section on the $QFT$. The circuit then applies the $QFT^\dag$ on qubit $a_0$, $a_1$ and $a_2$ and then measures those 3 qubits. This particular circuit estimates a generic T-gate. This example is outlined in [4] as well. A T-gate rotates the qubit by $\pi/4$, with a matrix of:\hfill

\begin{align*}
T = \begin{bmatrix}
1 & 0 \\
0 & e^\frac{i\pi}{4}
\end{bmatrix}
\end{align*}

\medskip
If we apply the T-gate to a qubit in the $\ket{1}$ state we get:

\begin{align*}
T\ket{1} = \begin{bmatrix}
1 & 0 \\
0 & e^\frac{i\pi}{4}
\end{bmatrix}
\begin{bmatrix}
0\\
1
\end{bmatrix}
= e^\frac{i\pi}{4}\ket{1}
\end{align*}

This means that we get get a phase applied to the qubit equal to $e^\frac{i\pi}{4}$. Since the generic phase of qubit is $e^{2i\pi\theta}$ we can say that $\theta$ is $\theta = \frac{1}{8}$. As result, when we execute the quantum phase estimation for the T-gate we should get a result of $\theta = \frac{1}{8}$. When we run this circuit we run it for a multiple shots, or measurements. For this circuit we will run it for shots = 1024.\hfill

\begin{figure}[htbp]
\begin{center}\vspace{0.2cm}
\includegraphics[scale=0.5]{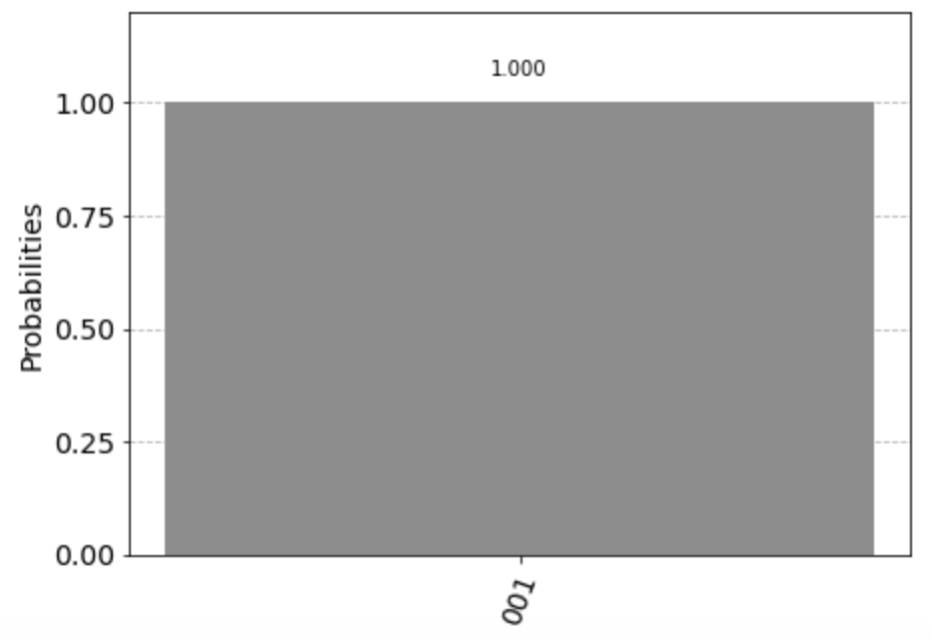}
\includegraphics[scale=0.5]{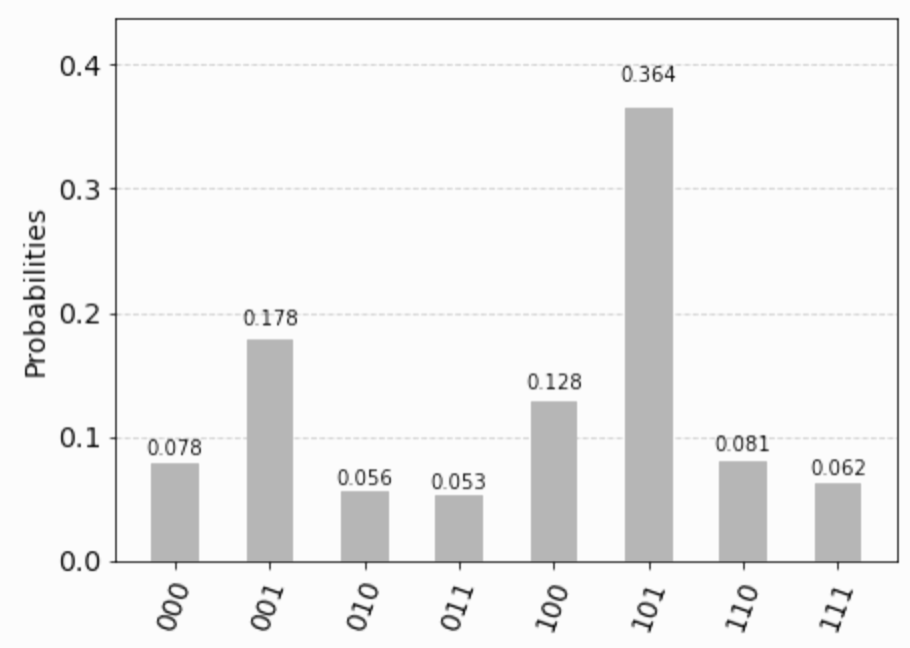}
\caption{Phase Estimation Results: (left) aer-simulator, (right) ibmq-manhattan}
\end{center}
\end{figure}
\FloatBarrier

\medskip
In figure 6(a) we see that there is 100 percent chance that we get the bit string '001'. However, if we rotate the qubit by an odd amount, such as $\pi/3$ we will get a less accurate phase estimations of the gate. As a result, there will be a certain percentage of states other the true value of the phase that are present. Here is where we take advantage of this phenomena to create a responsive music system using the physical properties of the quantum computer. Not only is the property of phase kickback utilized to create more flexible distribution of states within the phase domain but the noise from the real devices can be utilized to provide more variation in states represented. This is illustrated in figure 6(b) where the phase estimation circuit is run on both the simulator and on IBMQ Manhattan off the IBMQ net hub. We observe that we expect to see 100 percent of measuring the state '001'. This means that we are estimating the gate to apply the phase of
\begin{align*}
\theta = \frac{y_n}{2^n} = \frac{1}{2^3} = \frac{1}{8}\\
\end{align*}

In general, we not only have to consider the incorrect or mixed phases being applied we have to deal with the noise of these devices. As result, we will have to study how these two elements interact with one another.\hfill

\section*{PreProcessing and Mapping Audio Signals to Qubits}
\medskip
We want our system to use specific musical features from different sub-bands of our audio files to generate a new beat out of our samples from our database. To do this, we will take inspiration from work presented in [6] where a speech coding system is proposed using biological processes in the auditory nervous system (AN). In this work, speech signals are able represented using solely using the zero crossing metric from different sub-bands (i.e. low, mid and high frequency content) of the signals. For each sub-band a spike is used to represent a positive zero crossing event resulting in an sequence of impulses. This results, in a representation that requires low bit rate, and even though this compression algorithm is still lossy, the lost data is perceptually irrelevant. For QuiKo similar method will be implemented. A filter bank is applied to the input and database audio files creating filtered versions of each one. We will look at a simple case of dealing with just 3 sub-bands. Eventually, we will want to scale up to the full 25 sub-bands corresponding to the critical bands in the cochlea [7]. For now however, we will apply a low pass, band pass, and high pass filter to create three filtered versions of the signals. They will then be placed within a matrix to be encoded on the QuiKo Circuit which will be discussed in the next section.\hfill

\begin{figure}[htbp]
\begin{center}\vspace{0.2cm}
\includegraphics[width=1.0\linewidth]{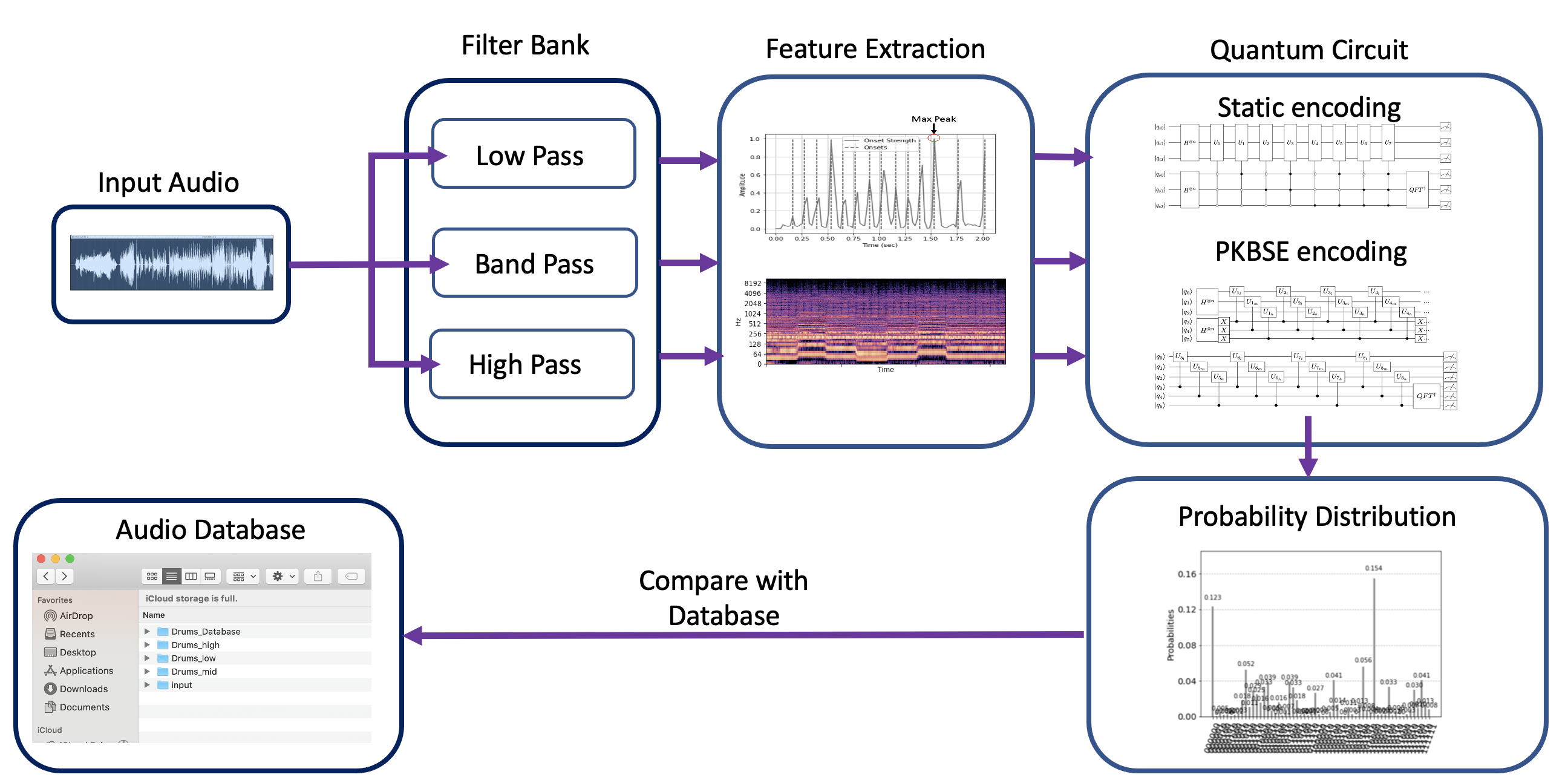}
\caption{QuiKo Architecture for Input Audio Signal}
\end{center}
\end{figure}
\FloatBarrier

\medskip
Music Producer Timbaland states "Everything is not a theory bruh...It's a feeling" [8]. As a result, the QuiKo methodology overall does not take on a rule based approached. It is based in the sonic content of audio samples being chosen and combined together. This application is focused on implementing an organic approach to generating music, attempting to give the system a sense of intuition, a "gut feeling". Quantum computing is well suited for this due to the fact that it can consider many different possibilities and outcomes simultaneously as do human musicians in the music creation process. This is the fundamental concept behind QuiKo's process for music generation in which we will call this approach Organic Rule based.\hfill

\subsection*{Drum Sample Database Preparation}
\medskip
First we need to prepare a database of audio samples to be used in the construction of the new generated beat. We will gather a collection of audio samples (i.e. single drum hits and long melodic and harmonic patterns and progressions). We then apply the filter bank as specified previously to each of the samples in the database. There should be a low, mid and high versions of each sample. For each of the samples’ filtered versions the Harmonic Percussive Source Separation (HPSS) algorithm from the librosa library in python [9] is then applies to extract harmonic and percussive features of the signals. The algorithm returns two signals via median filtering [9]. (1) percussive part where the transients and onsets of the signal are more pronounced (2) the harmonic part where the tonal and spectral content is more defined. These resulting signals are shown in figure 7. For the percussive part shown figure 7(a), the values of the peaks(spikes) in the signal are identified and are summed together. This sum is then divided by the value of the maximum peak, which will become our $\theta$ angle for the unitary gates used in the quantum circuit. The parameters and matrix for the U gate (U3 gate in qiskit) is expressed in equation (5).\hfill

\begin{figure}[htbp]
\begin{center}\vspace{0.2cm}
\includegraphics[width=0.9\linewidth]{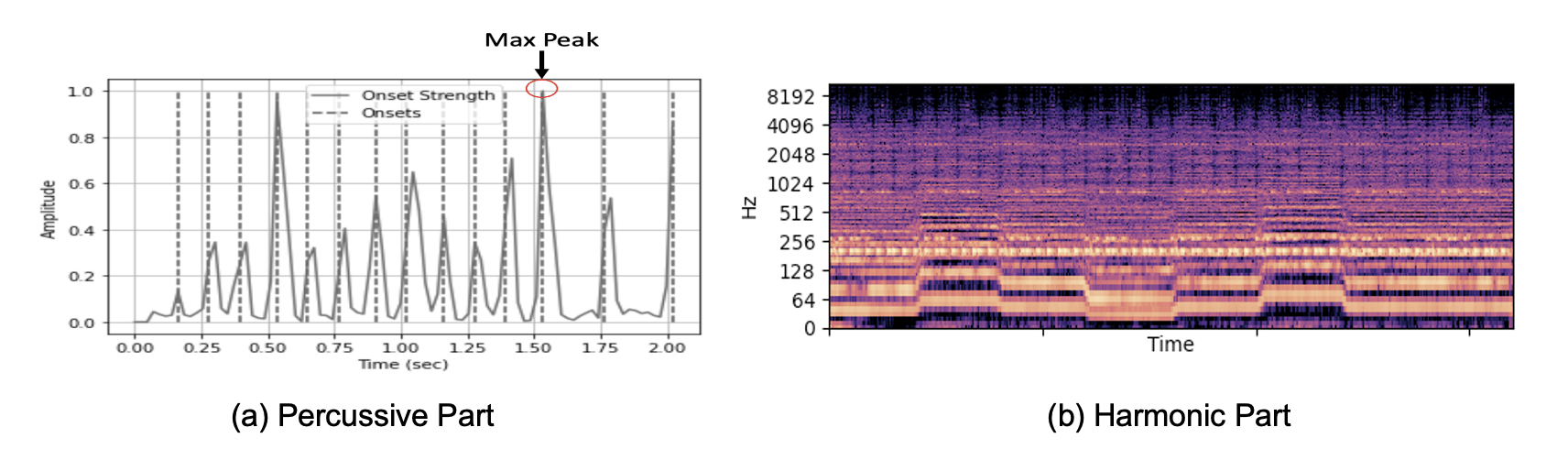}
\caption{Percussive and Harmonic parts of the Audio Signal}
\end{center}
\end{figure}
\FloatBarrier

\medskip
For the harmonic part of the signal, shown in figure 7(b), the Fast Fourier Transform (FFT) is performed. From there the highest 3 peaks are identified within the spectrum, and the weighted average of these values is calculated. This will be our $\phi$ parameter for the U-gates. Finally, the spectral centroid is also calculated from the the harmonic part which will define our $\lambda$ parameter.\hfill

\begin{figure}[htbp]
\begin{center}\vspace{0.2cm}
\includegraphics[width=0.7\linewidth]{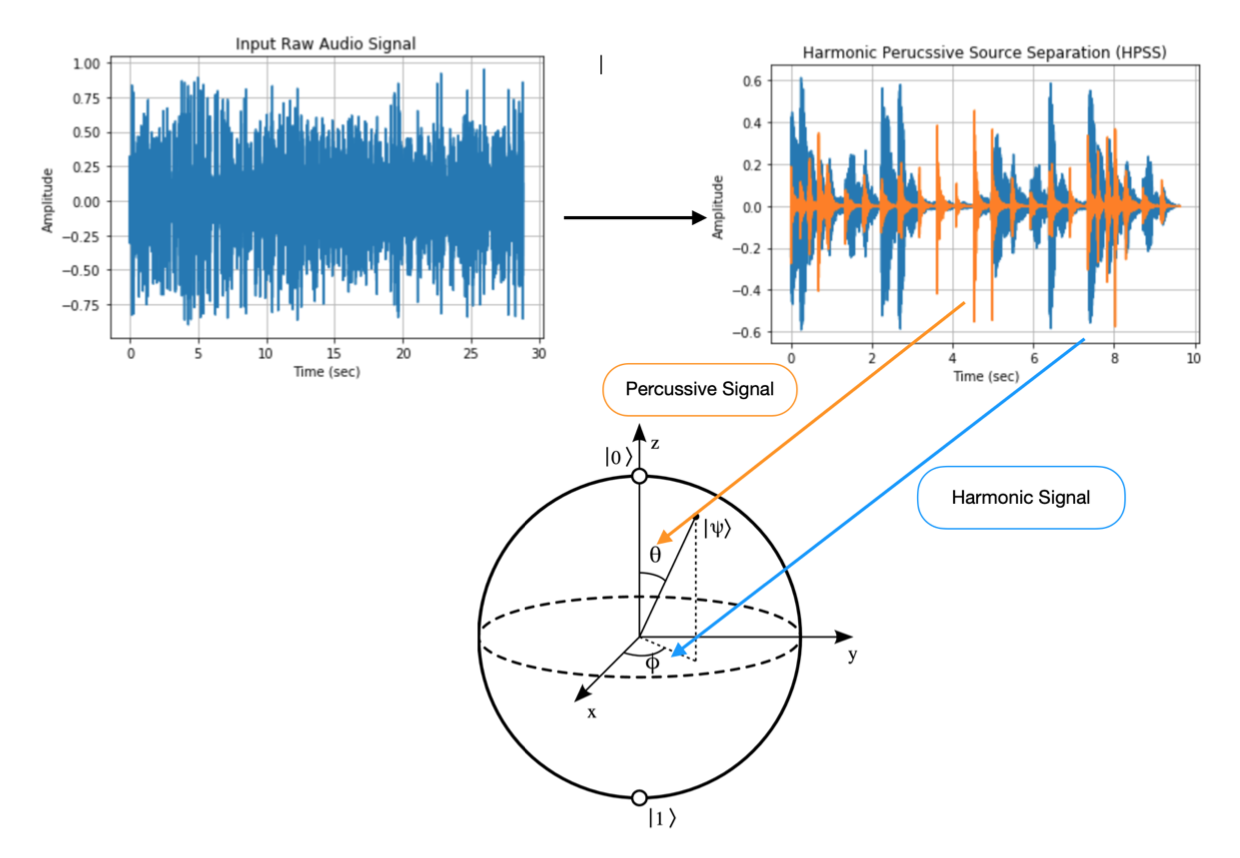}
\caption{HPSS Qubit Encoding}
\end{center}
\end{figure}
\FloatBarrier

\medskip
Equation (5) above expressed the U-gate operation in matrix form. Also it defines the parameters that are encoded onto each U-gate in the quantum circuit. Methods for this encoding will be discussed in further detail in the following sections. Also keep in mind that any set of features can be extracted and used as the parameter for these gates.\hfill

\begin{equation}
U(\theta, \phi, \lambda) = \begin{pmatrix}
\cos{\frac{\theta}{2}} & -e^i\lambda sin{\frac{\theta}{2}}\\
e^i\phi \sin{\frac{\theta}{2}} & e^i(\phi+\lambda) \cos{\frac{\theta}{2}}
\end{pmatrix}
\end{equation}

\begin{equation}
\phi = \frac{\sum_{n=0}^{N-1} f(n)x(n)}{\sum_{n=0}^{N-1} x(n)}\\
\end{equation}
\begin{equation}
\lambda = max \{ f(n)_{onset} \}\\
\end{equation}
\begin{equation}
\theta = arg max_{x = s} \{\sum_{n=0}^{N-1} x_n e^{\frac{-i2\pi kn}{N}} k = 0,...,N-1\}
\end{equation}

\newpage
\subsection*{Sample Database, Quantum Circuit \& States}
The calculations in the previous section will be done for each of the filtered versions of the original samples. The values in equations (6)(7)(8) will be encoded onto U-gates and applied to specific qubits. The angles calculated in (3) for the low version of the sample will be mapped to $q_0$ using a U3 gate in Qiskit[3]. The angles for the mid will be mapped to $q_1$, and high mapped to $q_2$.\hfill

\begin{figure}[htbp]
\begin{center}\vspace{0.2cm}
\includegraphics[width=1.0\linewidth]{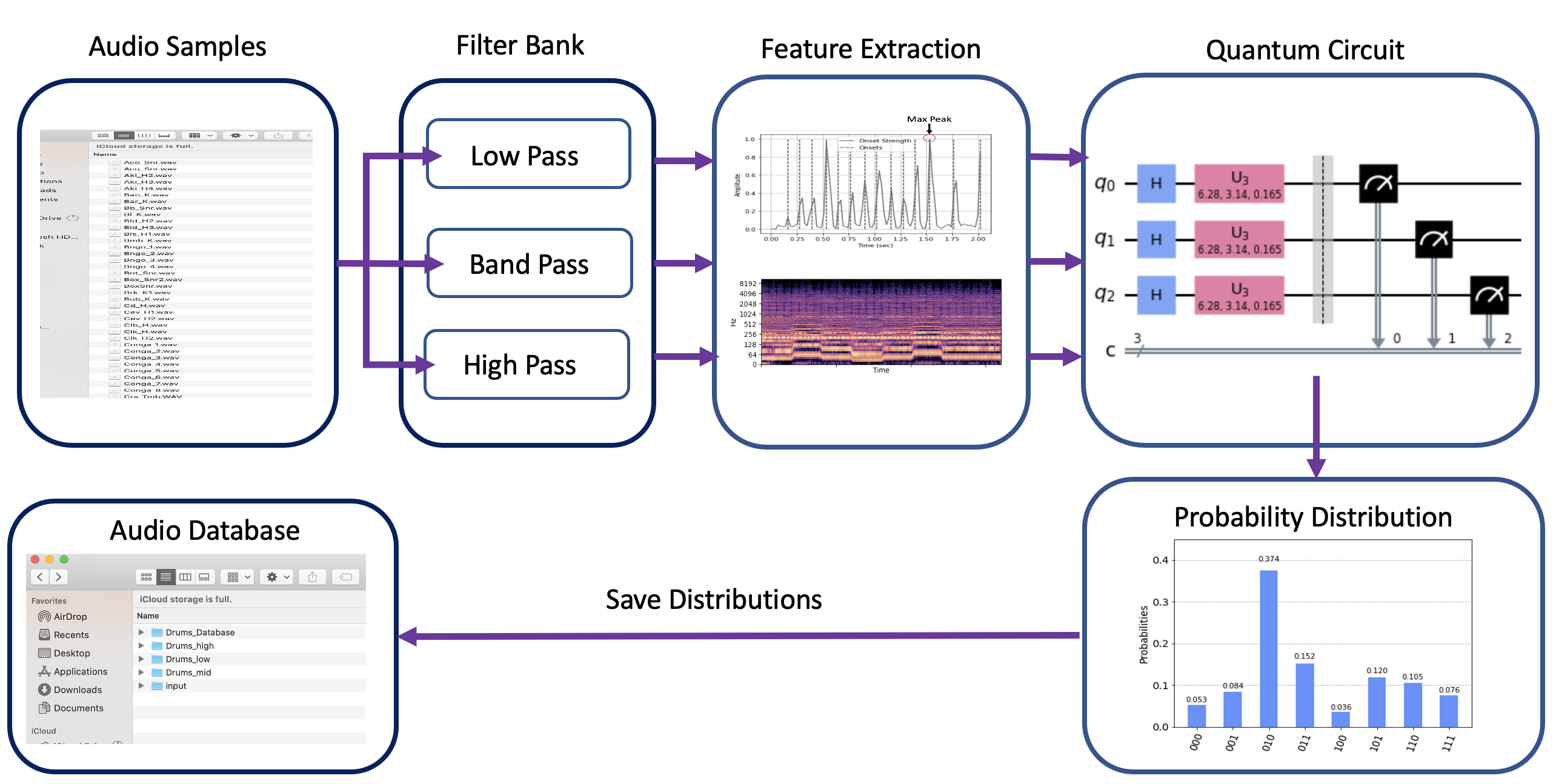}
\caption{HPSS Qubit Encoding}
\end{center}
\end{figure}
\FloatBarrier

\medskip
The circuit in figure 9 is then executed for 1024 shots. The resulting probability distribution for each audio track is then stored for future use. This process is repeated for each audio sample in the database. 

\section*{The Quantum Circuit (QuiKo Circuit)}
The process for preparing the database is similar to that of the input audio file that we want our output beat to be influenced by. The input audio file is filtered through the same filter bank that was used for the audio files in the database. So in this case we will get three filtered versions (low, mid and high bands) of the input audio file. Then, as we did for the database, we applied the HPSS algorithm to each filtered version getting two separate signals (Percussive part and the harmonic part) for each.

\medskip
The percussive and harmonic parts are then segmented into subdivisions depending on the number of qubits available in our circuit. Here we will allocate 3 qubits for our subdivision register in which we will call our spinal cord register. Since we have 3 qubits in our register we will divide the parts into 8 subdivisions corresponding to eight notes. For each subdivision between the first eighth note and the last eighth note we will apply a U-gate with the same feature set that we extracted from the database audio files. In other words, the averaged onset strengths of the percussive part of the input signal will map to $\theta$, the weighted average of the 3 highest frequency peaks in spectrum of the harmonic part of the input signal will map to $\phi$, and the spectral centroid of the harmonic part will be mapped to $\lambda$ of our U-gates for each subdivision. Again, this will be done for each filtered version of the input audio file. Once these features have been extracted for each subdivision of each filtered version of the signal and encoded them as parameters on our U-gates, we need to associate each U-gate with a specific a specific subdivision. The way this is done is through entangling another register of qubits, where wee will apply the encoded U-gates, to the spinal cord register. This will entangle a particular U-gate to its corresponding subdivision.\hfill

\medskip
This can be done in various ways. In this section we will discuss two methods of encoding these musical feature on to qubit registers and entangling them with their corresponding subdivision information. These methods include (1) Encoding Static (2) Phase Kickback Sequencing Encoding.\hfill

\subsection*{Static Encoding}
\medskip
This method is based off the QRDA[10] and FQRDA[11] quantum representations of audio signals. In general, the extracted musical features per sub-band are encoded on to the quantum circuit and is entangled to its corresponding subdivision. Breaking down figure 5 we see that the circuit is divided into two qubit registers timbre register and spinal cord register. We first prepare both qubit registers in equal superposition by applying a single Hadamard gate to each qubit in the circuit so that we have equal probability of getting each subdivision. All these initial gates on both registers are referred to as the internal pulse of the system. This is analogous to a musicians personalized sense of 'groove' or rhythm based on their past musical experiences. For now we will only deal with the equal superposition case as we want to see how the system will perform with equal probability of getting each eight note subdivsion.\hfill

\medskip
Next we set up a series of cascading multi-controlled Unitary gates. Each of these U3 gates are applied depending on the subdivision that the spinal cord register collapses to. Note that the controls represented as closed black filled dots are checking to see if the qubit happens to collapse to '1', and the controls represented as empty dots are checking to see if the qubit collapses to '0'. For example, in figure 5 the multi-controlled U-gate U5 has a closed black filled control on the first and the third qubits, and a empty control on the second qubit in spinal cord register. This means that the U-gate U5 will be applied to the timbre register if the spinal cord register collapsed to $\ket{101}$, or the 5th subdivision in the measure.\hfill

\begin{figure}[htbp]
\begin{center}\vspace{0.1cm}
\includegraphics[scale=0.4]{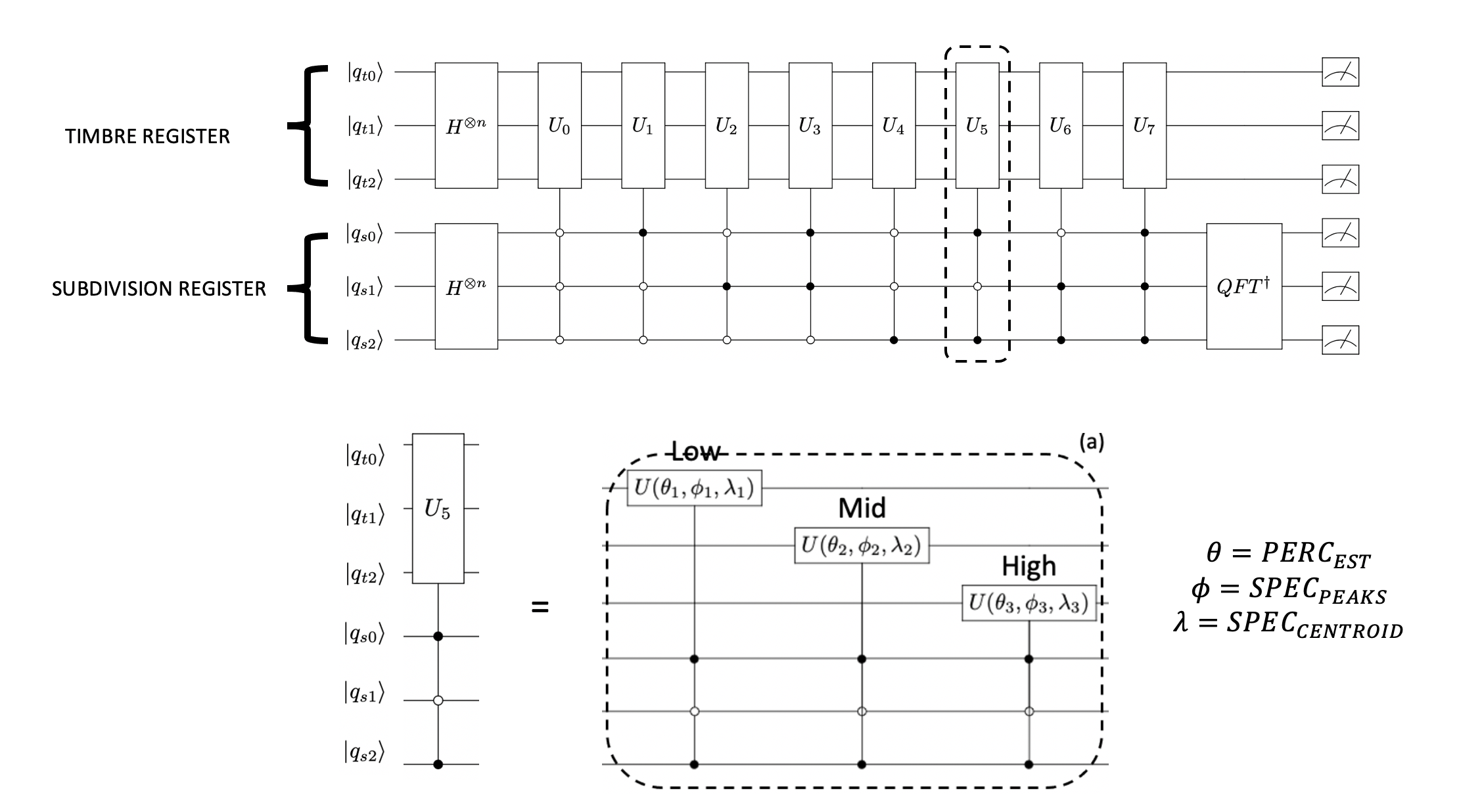}
\caption{Static Encoding Circuit}
\end{center}
\end{figure}
\FloatBarrier

\medskip
Each of the multi-controlled U-gates in figure 5 contain three separate multi-controlled U3 gates. Each corresponding for a different sub-band on a particular subdivision. We can also see that for each gate on each sub-band we see the parameters associated with the musical feature we extracted for a corresponding subdivision. Qubit $q_0$ is allocated for parameters in the low band, $q_1$ is allocated for parameters in the mid band, and $q_2$ is allocated for parameters in the high band. As a result, the timbre register will collapse to a 3 bit binary string, and thus when we measure it many times we get a probability distribution associated with a particular subdivision.\hfill

\begin{figure}[!htb]
\begin{center}\vspace{0.1cm}
\includegraphics[scale=0.75]{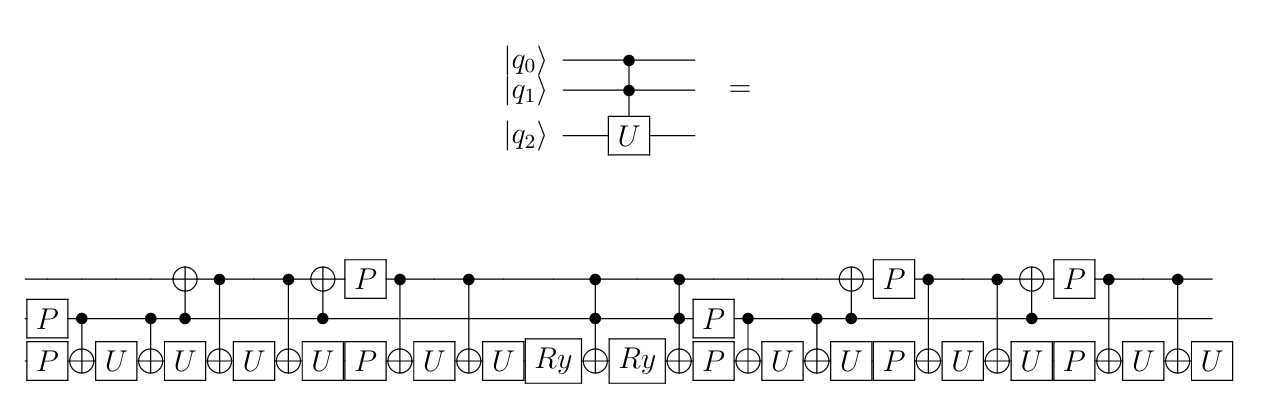}
\caption{Multi-Controlled U-gate Decomposition}
\end{center}
\end{figure}

\medskip
As each of these multi-controlled U-gates are applied to the timbre register, depending on the collapsed state of the spinal cord register, the phase of the corresponding U-gate is kicked back to the spinal cord register. So if we consider the case of $U_5$ again, the phase associated with those set of gates will be pushed into the spinal cord register thus changing is state in the Fourier basis. In other words, the state it was upon triggering the $U_5$ is now offset in the Fourier basis. Thus, if we measure the spinal cord in the Fourier basis we will obtain a different subdivision than that the resulting timbre code was originally entangled with. To do this phase estimation is performed on the spinal cord register by applying the $QFT^\dagger$ to the spinal cord register and then measure it.\hfill

\subsection*{Phase Kickback Sequencing}
\medskip
Static Encoding, however, is very expensive to implement as multi-controlled qubit gates (containing more than one control) do not correspond to the cost of the number of controls and targets. For example, a controlled X (cx) gate would have the cost of 2 for target and the control qubits [12]. Any more than one control qubit would need to decompose into a larger circuit as shown for the multi-controlled U-gate with 2 controls in figure 11.As a result, if we want to design our algorithms to deal with subdivisions any small than eight notes, the circuit cost would drastically increase. Alternative, more cheaper methods are needed if we want to scale our system for more detail and resolution.\hfill 

\medskip
Here we propose a method in order to reduce the cost associated with static encoding. This method is called Phase Kickback Sequence Encoding (PKBSE). In previous sections we discussed the effects of phase kickback produced by controlled quantum gates and how to estimate the amount of phase that gets kicked back into the control qubit register (in this case the spinal cord register).In order to reduce the cost of the static encoding circuit we need to replace the multi-controlled U-gates with single controlled U-gates, and sequence them in a way that apply parameters of a specific subdivision. Figure 12 outlines the logic behind this method. This has a number of steps:\hfill

\begin{enumerate}
   \item Split the measure in half with half the subdivisions on one side (starting with 0 in binary) and the other on the right side (subdivisions starting with '1' in binary).
   \item Calculate and/or extract the desired feature for each subdivisions on left and right side of the measure.
   \item For one side of the measure (the '0' side or the '1' side) sum together the features associated with each subdivision with the same type of features in previous subdivisions. This is done to reflect the behavior of a human musician in which their musical response is based on the current and previous musical events from other performing entities.
   \item multiply all the feature summations by -1 if they are not the associated with the final subdivision for each half of the measure.
   \item repeat this process for the other side of the measure.
   \item Organize the data into a PKBSE encoding matrix as at the bottom of figure 12. 
\end{enumerate}

\begin{figure}[htbp]
\begin{center}\vspace{0.1cm}\hspace{0.1cm}
\includegraphics[scale=0.45]{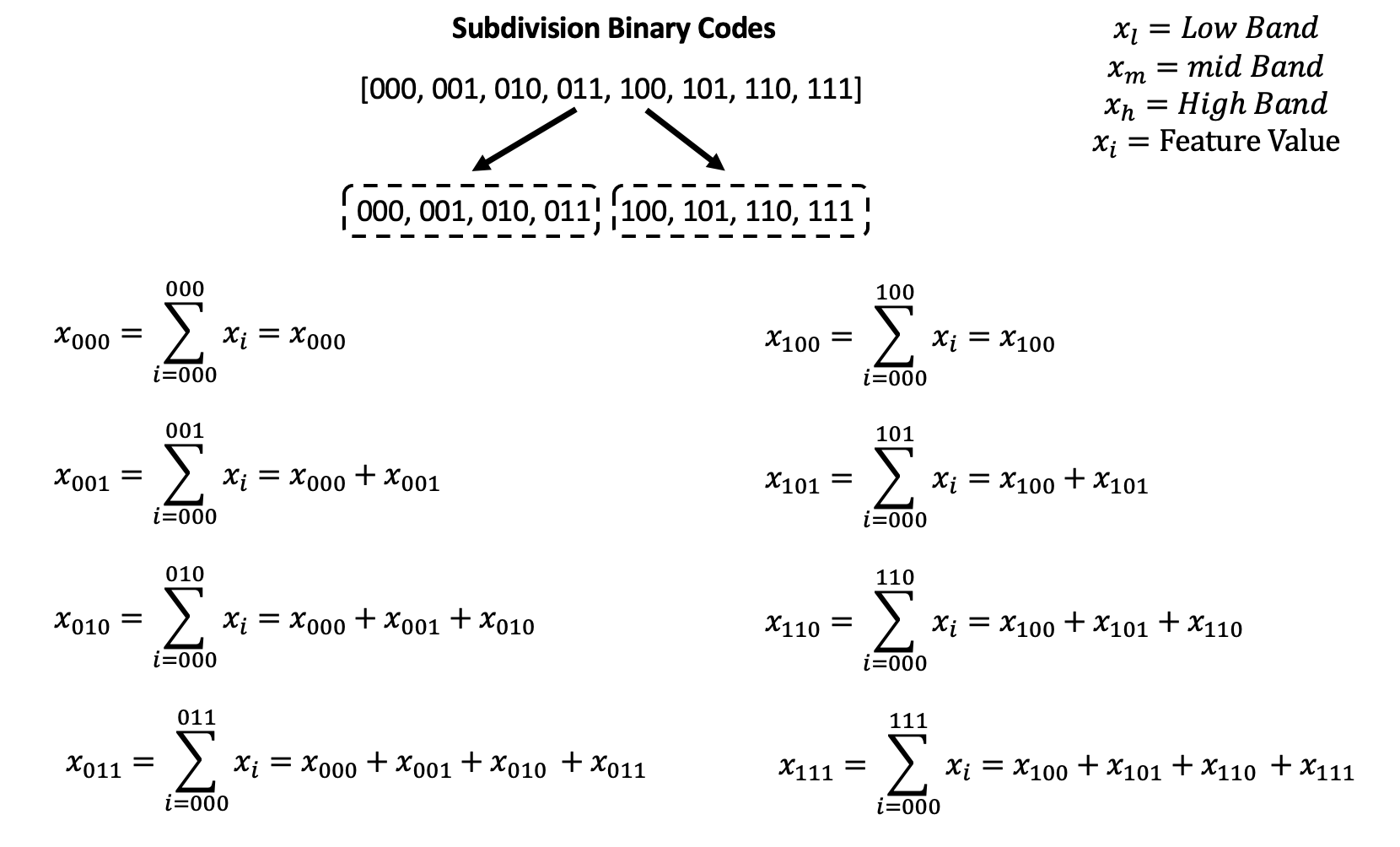}
\includegraphics[scale=0.45]{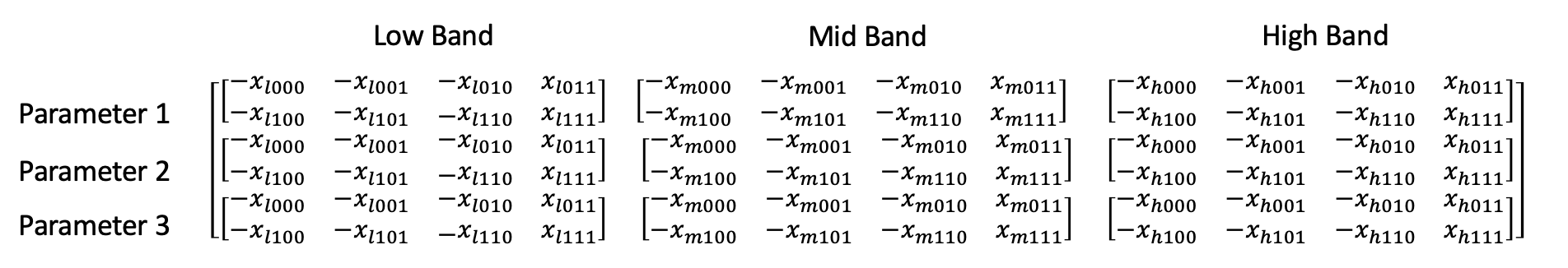}
\caption{PKBSE Encoding Representation}
\end{center}
\end{figure}
\FloatBarrier

\medskip
We negate all summed other than the last subdivision within the respective halves of the measure due to the fact that they cover the entire segment. If we sum all the parts together for a particular sub-band we get a sequence dependent on the qubits themselves being 0 or 1 and remove the smaller segments from the total feature value from the higher segment layers. After we have done this for each feature we organize it into an encoding matrix shown at the bottom of figure 6 in order to organize our parameters to encode onto our quantum circuit. Since we are dealing with a low, mid and high band along with 3 separate parameters, our PKBSE encoding matrix will be 3x3. Each element this matrix will be 2x4 matrix containing each of the summed features for each subdivision.\hfill

\begin{align*}
\Qcircuit @C=0.03em @R=0.0025em {
&\lstick{\ket{q_0}} & \multigate{2}{H^{\otimes{n}}} & \qw & \gate{U_{{1}_{l}}} & \qw & \qw & \gate{U_{{2}_{l}}} & \qw & \qw & \qw & \gate{U_{{3}_{l}}} & \qw & \qw & \qw & \qw & \gate{U_{{4}_{l}}} & \qw & \qw & \qw && ...\\
&\lstick{\ket{q_1}} & \ghost{H^{\otimes{n}}} & \qw & \qw & \gate{U_{{1}_{m}}} & \qw & \qw & \gate{U_{{2}_{l}}} & \qw  & \qw & \qw & \gate{U_{{3}_{m}}} & \qw  & \qw & \qw & \qw & \gate{U_{{4}_{m}}} & \qw & \qw && ...\\
&\lstick{\ket{q_2}} & \ghost{H^{\otimes{n}}} & \qw & \qw & \qw & \gate{U_{{1}_{h}}} & \qw & \qw  & \gate{U_{{2}_{h}}} & \qw & \qw & \qw & \gate{U_{{3}_{h}}} & \qw & \qw & \qw & \qw & \gate{U_{{4}_{h}}} & \qw && ...\\
&\lstick{\ket{q_3}} & \multigate{2}{H^{\otimes{n}}} & \gate{X} & \ctrl{-3} & \qw & \qw & \ctrl{-3} & \qw & \qw & \qw & \ctrl{-3} & \qw & \qw & \qw & \qw & \ctrl{-3} & \qw & \qw & \gate{X} && ...\\
&\lstick{\ket{q_4}} & \ghost{H^{\otimes{n}}} & \gate{X} & \qw & \ctrl{-3} & \qw & \qw & \ctrl{-3} & \qw & \qw & \qw & \ctrl{-3} & \qw & \qw & \qw & \qw & \ctrl{-3} & \qw & \gate{X} &\qw & ...\\
&\lstick{\ket{q_5}} & \ghost{H^{\otimes{n}}} & \gate{X} & \qw & \qw & \ctrl{-3} & \qw & \qw & \ctrl{-3} & \qw & \qw & \qw & \ctrl{-3} & \qw & \qw & \qw & \qw & \ctrl{-3} & \gate{X} & \qw & ...
}
\end{align*}

\begin{align*}
\Qcircuit @C=0.5em @R=0.0025em {
&\lstick{\ket{q_0}}  & \gate{U_{{5}_{l}}} & \qw & \qw & \gate{U_{{6}_{l}}} & \qw & \qw & \qw & \gate{U_{{7}_{l}}} & \qw & \qw & \qw & \qw & \gate{U_{{8}_{l}}} & \qw & \qw & \qw & \meter\\
&\lstick{\ket{q_1}}  & \qw & \gate{U_{{5}_{m}}} & \qw & \qw & \gate{U_{{6}_{m}}} & \qw  & \qw & \qw & \gate{U_{{7}_{m}}} & \qw  & \qw & \qw & \qw & \gate{U_{{8}_{m}}} & \qw & \qw & \meter\\
&\lstick{\ket{q_2}} & \qw & \qw & \gate{U_{{5}_{h}}} & \qw & \qw  & \gate{U_{{6}_{h}}} & \qw & \qw & \qw & \gate{U_{{7}_{h}}} & \qw & \qw & \qw & \qw & \gate{U_{{8}_{h}}} & \qw & \meter\\
&\lstick{\ket{q_3}} & \ctrl{-3} & \qw & \qw & \ctrl{-3} & \qw & \qw & \qw & \ctrl{-3} & \qw & \qw & \qw & \qw & \ctrl{-3} & \qw & \qw & \multigate{2}{QFT^\dagger} & \meter\\
&\lstick{\ket{q_4}} &  \qw & \ctrl{-3} & \qw & \qw & \ctrl{-3} & \qw & \qw & \qw & \ctrl{-3} & \qw & \qw & \qw & \qw & \ctrl{-3} & \qw & \ghost{QFT^\dagger} & \meter\\
&\lstick{\ket{q_5}} &  \qw & \qw & \ctrl{-3} & \qw & \qw & \ctrl{-3} & \qw & \qw & \qw & \ctrl{-3} & \qw & \qw & \qw & \qw & \ctrl{-3} & \ghost{QFT^\dagger} & \meter
}
\end{align*}

\medskip
Figure 7 shows the quantum circuit for the PKBSE encoding method. The spinal cord and timbre registers are setup in the same way that they were in static encoding. Each qubit in the timbre register represents one of the sub-band of the input audio signal, while the spinal cord register represent the different subdivisions that are being considered. This is done by entangling the two registers in a particular way. We will use the concept presented in [13] which states that human musicians perceive the attack times of instrument with lower frequency content with less resolution than that of instruments with high frequency content. Here we can say that parameters associated with the low sub-band, encoded on to $q_0$, will be entangled with the most significant qubit in the spinal cord register, $q_3$. This is due to the fact that the rate at which $q_3$ changes is less frequent that the other qubits in the register. Following suit, $q_1$ which deals with mid sub-band sequence parameters will be entangled with the next significant qubit $q_4$, and so on and so forth.\hfill

\medskip
The separation between the sequences for the first and the second half of the measure can be observed in the circuit as well. The first half of the measure (as stated previously) is defined by '0' in the most significant spinal cord qubit, and thus its U-gate sequence is enclosed by X gates on the spinal cord register. This sequence of gates will be triggered if any of the spinal cord qubits happen to be '0'. On the other hands, if any of these qubits happen to be '1' then the gate sequence outside of the X gates will be triggered. The encoding process of mapping the extracted features of the signal to parameters on their corresponding controlled U-gates is identical to that for static encoding. However, in the PKBSE circuit we will get a direct phase kick back from the U-gates that were applied to the timbre register, and thus elements from the original signal should have more direct impact on the states for the spinal cord register. Also in contrast to the static encoding method where we considered the effects of features for one subdivision at a time, the PKBSE method allows the system to consider the effects of groups of subdivisions at the same time in superposition.\hfill

\section*{Results}
\subsection{Decoding \& Beat Construction}
\medskip
Once the core quantum algorithm has been executed on either a simulator or real device, we want to decode the results in order to construct the final beat. To do this we need to compare the probability distributions generated by the executing the quantum circuits for our audio files for each subdivision of the input signal. To compare the different these quantum states the fidelity between the state of the input track and the database tracks are calculated. Fidelity measures how close two quantum states are to each other [14], and thus will identify which audio files in database are most (and least) similar to the input audio file.\hfill

\begin{equation}
    F( \rho,\sigma ) = ( tr(\sqrt{\rho}\sigma \sqrt{\rho}) )^2 = | \braket{\psi_{\rho} | \psi_{\sigma}} |^2
\end{equation}

\medskip
After the fidelity is calculated for all the database and the input audio files, the audio samples in the database are organized into layers based on the value of each sample's fidelity. A layer is a group of audio samples that occupy a single subdivision.\hfill

\begin{figure}[htbp]
\begin{center}\vspace{0.1cm}
\includegraphics[scale=0.55]{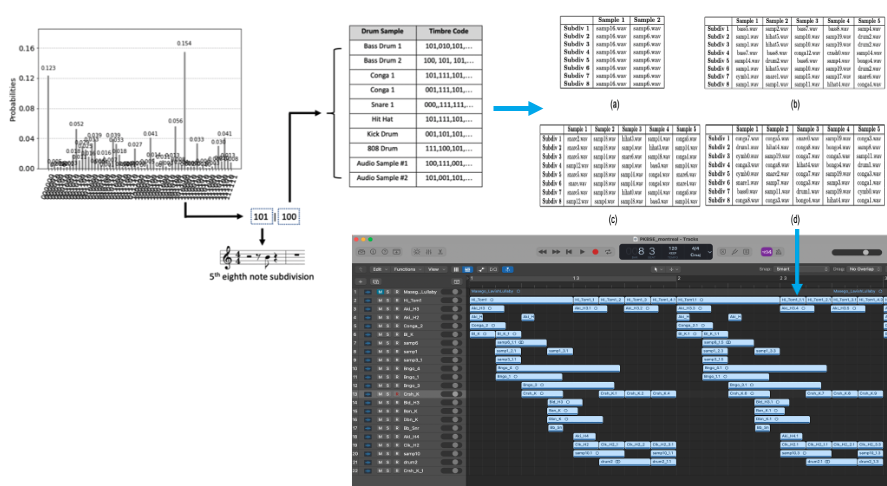}
\caption{Results for Spinal Cord Qubit Register}
\end{center}
\end{figure}
\FloatBarrier
\medskip
After some experimentation and listening, it was found that high fidelity values led to more pulsating and repeating audio sample sequences. Layers further away from the input audio signal begin to present more rhythmic patterns and sequences with more variation. An example of this is illustrated in figure 13. There is a trade off between consistent spectral content and the rhythmic variation to the input audio signal.\hfill

\begin{figure}[htbp]
\begin{center}\vspace{0.1cm}
\includegraphics[scale=0.43]{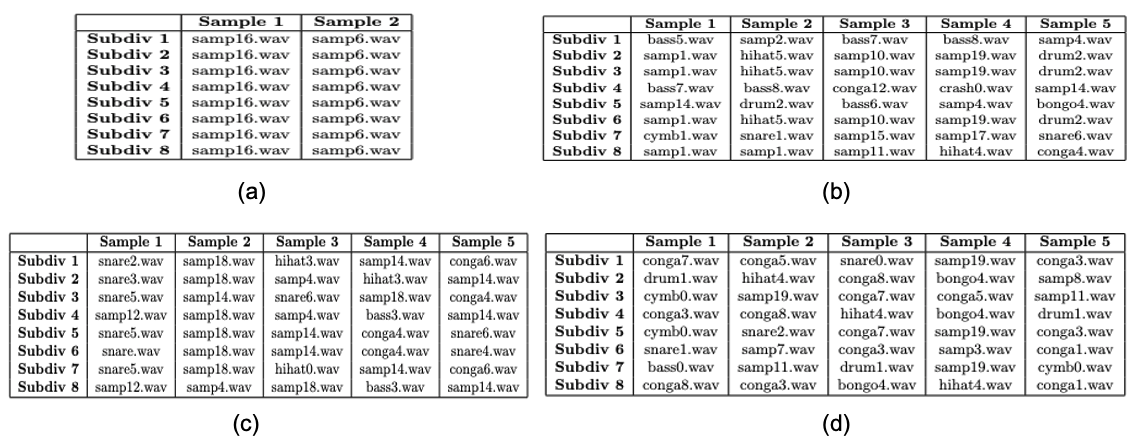}
\caption{Sample Results of Static \& PKBSE Encoding Decoded Layer Tables (a) Static Encoding maximum fidelity (b) Static encoding minimum fidelity (c) PKBSE encoding maximum fidelity (d) PKBSE encoding minimum fidelity}
\end{center}
\end{figure}
\FloatBarrier
\subsection*{Analysis}
\medskip
What does this mean in terms of how the system performs? How do the two encoding methods we discussed previously (Static and PKBSE encoding) compare with one another? To get a better idea of how our system behaves for a wide variety of input signals a matrix of random parameters is used as input to the system for a number of trials. To compare the performance these two methods we will look at their flexibility in choosing different audio samples from the database. In addition, we must also measure the impact that phase kickback and noise associated with real quantum devices has on the newly generated beat patterns. Thus, we will need to utilize the expressibility measure proposed in [15]. This measure is primarily used to indicate the degree that a quantum circuit can explore Hilbert space. However, in this case we will adapt it to measure how well our Static and PKBSE quantum circuits can explored our audio database. We will take the following steps:\hfill

\begin{enumerate}
   \item Generate a matrix of random parameters values to encode on to the quantum circuits used to generate the probability distributions associated with audio files within the audio sample database. 
   \item Generate a uniformly random matrix and encode it onto our quantum circuit for Static and PKBSE methods.
   \item For each subdivision calculate the fidelity between the resulting state and of the input and the states of the audio tracks in the database.
   \item Repeat to collect M samples. For this experiment we used 50 samples (M = 50).
   \begin{enumerate}
     \item Record the probability distribution of how often a particular layer occurs (layer with identical samples in it).
    \end{enumerate}
    \item After the distribution of the layer occurrence is generated, generate a uniform distribution of the layers.
    \item Repeat this process for each layer of samples.
\end{enumerate}

\medskip
Figure 15 plots the expressibility curve for both the static and PKBSE methods, executed on the simulator and IBMQ Manhattan. The x-axis shows the layer number, while the y-axis shows the expressibility value for the layer, while the plot on the left depicts the expressibility running the quantum circuits on qiskit's aer-simulator. The plot on the right depicts the expressibility results after running the quantum circuits on the real device from the IBMQ net hub. Here we are specifically using IBMQ Manhattan. Looking at these graphs we can see that the overall expressibility for the simulator is high for the static encoding method in comparison for for the lower layers. The higher the expressibility value the less it can explore the database for a variety of inputs.\hfill 

\begin{figure}[htbp]
\begin{center}\vspace{0.1cm}
\includegraphics[scale=0.575]{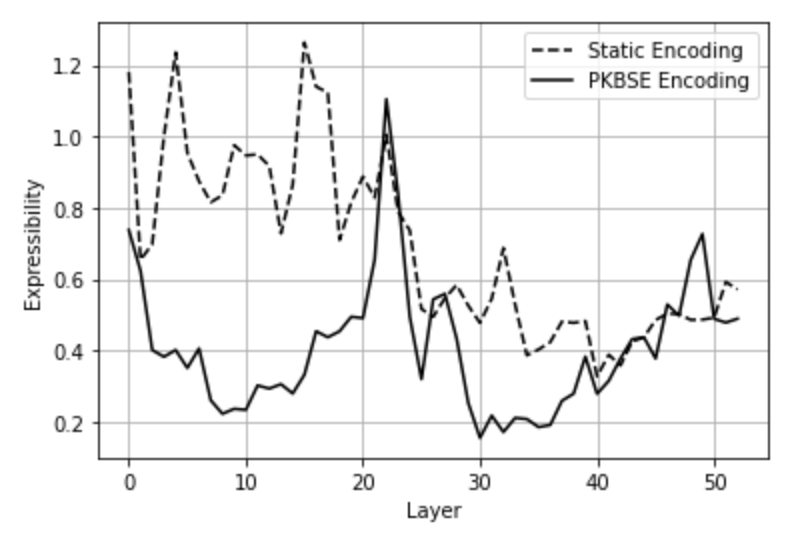}
\includegraphics[scale=0.575]{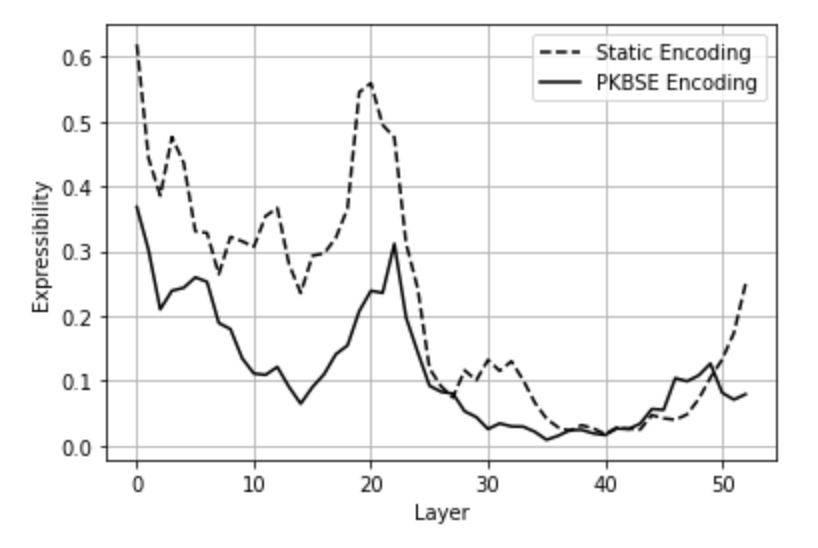}
\caption{PKBSE Encoding Circuit: aer-simulator(Right), Real Device(Left)}
\end{center}
\end{figure}
\FloatBarrier

\medskip
For the results obtained from running on the aer simulator it is observed that the lowest layer has a $70\%$ higher expressibility that the PKBSE. As the layer number increases the PKBSE decreases to a local minimum around layer 10. A spike in the expressibility curve then occurs between layers 20 and 25, approximately matching the expressibility value of the static encoding. We then see another local minimum at layer 30, with expressibility of approximately 0.2. After this, the curve begins to increase again starting at layer 35 and the static and PKBSE expressibility begin to converge. However, for static encoding the local minimums are not as pronounces as they are for the PKBSE method. There is more of a gradual decline for the static encoding method with oscillations about the general shape of the curve. The two expressibility curves for the static and PKBSE encoding then begin to converge with each other after layer 40.\hfill

\medskip
For results obtained from running on IBMQ Manhattan, both curves take on a gradual declining shape with a pronounced spike around layer 20. Here a more noticeable difference can be observed between the spike of the static and PKBSE expressibility curves. These spikes are also offset from one another by a layer. The curves then begin to converge to very low expressibility values until they diverge again after layer 40. This shape shows that the noise introduced by the real device lowers the expressibility value and in the case of the static encoding smooths out the curve. The oscillations associated with the static encoding method are now minimized and begins to look similar to the shape of the PKBSE curve. In contrast, the PKBSE expressibility curve maintains the same shape that was observed from the simulator. The noise introduced from the real quantum device scales the PKBSE curve down by a factor of 2 (approximately).\hfill

\begin{figure}[htbp]
\begin{center}\vspace{0.1cm}
\includegraphics[scale=0.55]{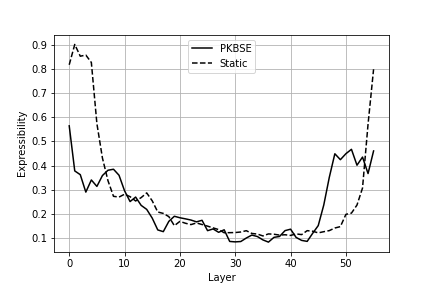}
\includegraphics[scale=0.55]{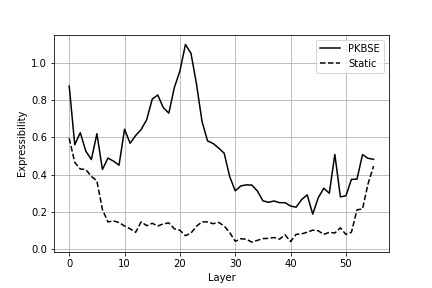}
\caption{PKBSE Encoding Circuit: aer-simulator(Right), Real Device(Left)}
\end{center}
\end{figure}
\FloatBarrier

\medskip
What we can conclude is that static and PKBSE encoding theoretically behave differently for various input values for a single database of audio samples. However, with the noise introduced by the real devices we see that they then begin to behave more similarly. In addition, it can also be concluded from analyzing these plots that the layers with the highest expressibility (least flexibility) for a randomized database are lowest, the highest and the layers half way between the highest and lowest layers. Figure 16 shows the expressibility curves of the system for both static and PKBSE circuits for a real audio sample database (non-randomized). When executed on the simulator, the results obtained are in line with what we found for the the randomized database run on IBMQ Manhattan with the exception that no spike within the mid-layers occurred for either method. Overall, for this database it is expected that the PKBSE has a lower expressibility (more flexibility) than the static encoding. The Static encoding however, has steeper slopes near the ends of the curves allowing for more flexibility with more of the inner layers. At the time of running the system for the results in figure 16 IBMQ Manhattan has been retired and all circuits needed to be run on a different device, IBMQ Toronto. The Static encoding expressibility curve for this database on IBMQ Toronto keeps it's same shape as seen for running on the simulator. But the expressibility curve for the PKBSE shows a massive spike, surpassing a value of 1.0 at layer 20, and spanning layers between 10 and 30. Thus, what has been observed is that the noise from the real devices can cause the expressibility curves to smooth out, scale down or scale up from from the shape of the expected results. As result, various types of databases with audio samples varying in timbres and spectral content need to be further studied.\hfill

\medskip
When encoding the musical information on to quantum circuits the perfect reconstruction of the musical data is not the primary concern. We can prepare the state of the qubit register so that different voices of the musical information can be generalized and operated on as single object. When a musician is improvising in the moment they are less concerned with the transcription of the music but rather how to react. So when the system is measured without any additional operations applied, it should produce a very flexible but still related interpretation of the original musical content, rather than replicating it.\hfill

\subsection*{Phase Kick Back Results \& Analysis}
\medskip
The expressibility metric primarily considers only the timbre and spectral aspects of the audio signals. However, we also need to analyze the other critical element of our system, phase kickback. As state previously, phase kickback contributes to the rhythmic response of the system. To analyze it we need to look at the effects that phase kick back has on the spinal cord register of our quantum circuit. We will follow a similar method as we did with the expressibility metric. We will take 50 samples of randomly generated parameters for both encoding methods, and then will obtain averaged probability distributions for the spinal cord qubit register when the circuits are executed. The results will then be compared to the a distribution showing an equal superposition of each eighth note subdivision. This will be done by computing the Kullback-Leibler Divergence (KLD) [16] between the averaged distributions of each of the encoding methods against the equal superposition produced by the quantum circuit in figure 17.\hfill

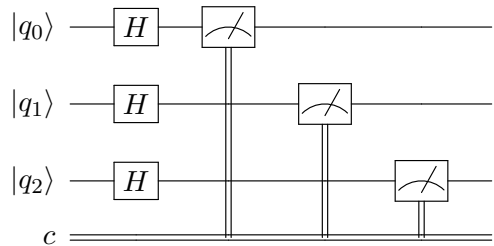
\begin{figure}[htbp]
\begin{align*}
\Qcircuit @C=1.5em @R=1.25em {
&\lstick{\ket{q_0}} &\gate{H} &\meter & \qw & \qw & \qw\\
&\lstick{\ket{q_1}} &\gate{H} & \qw & \meter & \qw & \qw\\
&\lstick{\ket{q_2}} &\gate{H} & \qw & \qw & \meter & \qw\\
&\lstick{c} & \cw & \cw \cwx[-3] & \cw \cwx[-2] & \cw \cwx[-1] & \cw
}
\end{align*}
\caption{Quantum Circuit for equal Superposition for 3 qubits}
\end{figure}

\medskip
Figure 18 shows the results for the spinal cord register for both the simulator and IBMQ Manhattan quantum computer for both the Static and PKBSE encoding methods. Distributions for the circuit shown in figure 8 are included to compare and observe the impact that phase kick back and noise introduced from the real device had on the results for both encoding methods. Let's first take a look at the simulator. In the upper left hand corner of figure 9 we see the distribution for equal superposition executed on the simulator. The distribution on the upper center of the figure shows the results for static encoding circuit, which produced a decrease in the probability that subdivision '000' or '111' would occur. It shifted energy from these subdivision to subdivisions '001', '011', '100' and '110', while '010' and '101' stayed the same. These are similar results observed for the PKBSE method.

\begin{figure}[htbp]
\begin{center}\vspace{0.1cm}
\includegraphics[scale=0.475]{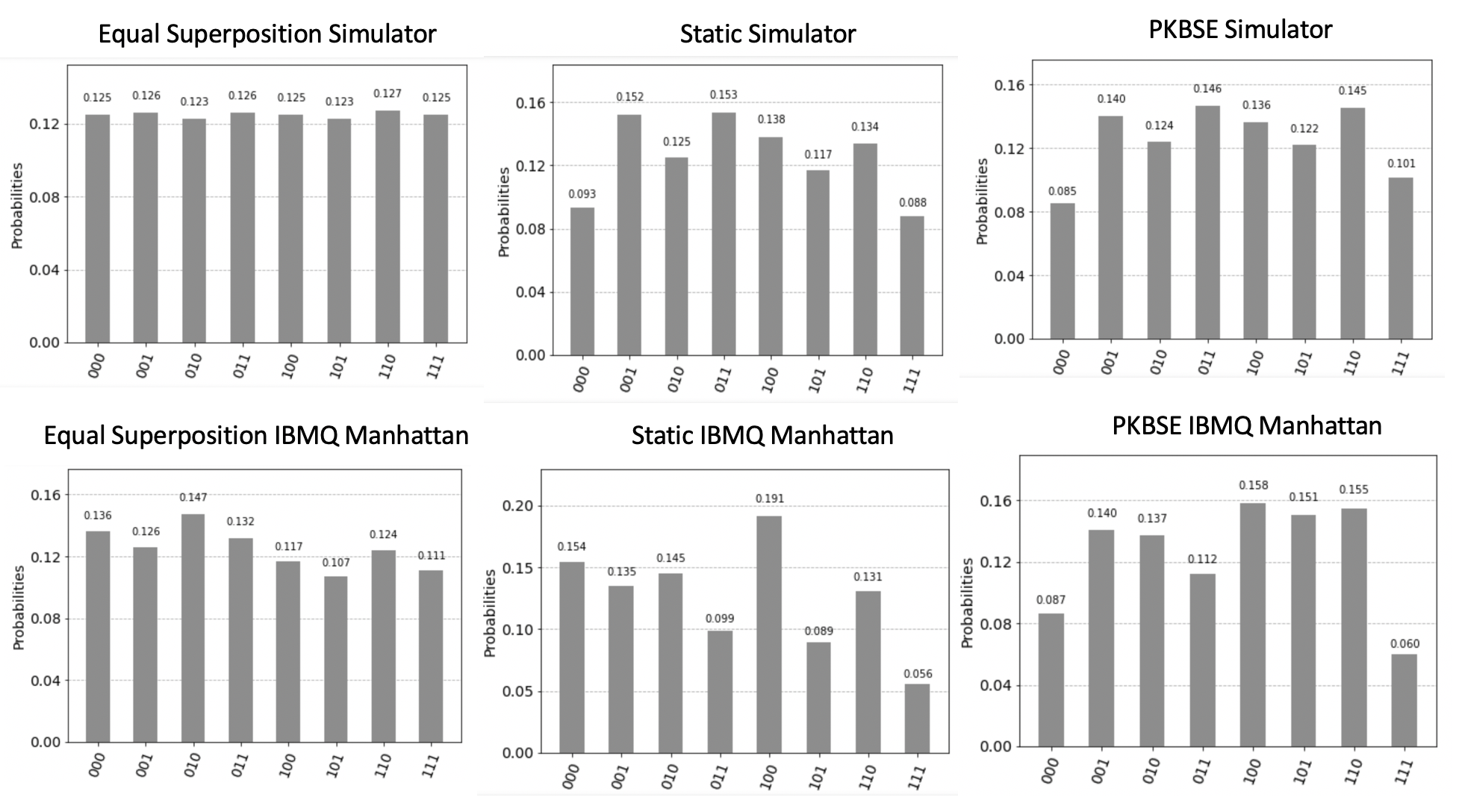}
\caption{Results for Spinal Cord Qubit Register}
\end{center}
\end{figure}
\FloatBarrier

\medskip
If we look at the results from the real device, we see that the static and the PKBSE averaged distributions for the spinal cord registers are now different. The phase introduced by the static encoding circuit on the spinal cord register caused the results to slightly skew right. The median shifts from '011' (subdivision 4), as seen in the simulator, to '010' (subdivision 3). This causes the first three subdivisions to increase its probability of occurring, with the except of '100' (subdivision 5), which has the highest probability to be selected within the measure. Comparing the KLDs calculated (table 3) for the simulator and IBMQ Manhattan for the static encoding, the KLD for the simulator case is $38.6\%$ smaller than KDL the real device case. This means that the phase kick back and noise associated with IBMQ Manhattan had a greater impact than expected from the simulator.\hfill

\begin{figure}[htbp]
\begin{center}\vspace{0.1cm}
\includegraphics[scale=0.75]{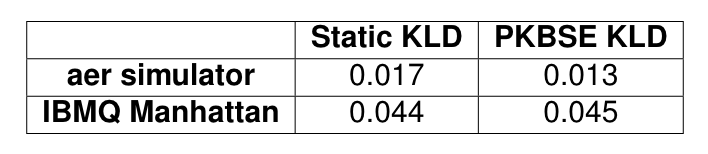}
\caption{KLD Table}
\end{center}
\end{figure}

\medskip
For the PKBSE there is a decrease in the right and left ends of the distribution in comparison to the equal superposition case for results obtained from the simulator and IBMQ Manhattan. However, the results for the real device are more consistent among groups of subdivisions. There is a decrease in amplitude at '011' (subdivision 4) causing the the distribution to take on a bi-modal shape, with a median of '100' (subdivision 5). The three most likely subdivisions that the PKBSE will select occur on the left side of the measure at '100' (subdivision 5), '101' (subdivision 6) and '110' (subdivision 7). For the right side of the measure, PKBSE will more likely choose '001' (subdivision 2) and '010' (subdivision 3). The KLD values for the PKBSE are also shown in table 3 and are very similar to the values for the Static encoding method.\hfill

\medskip
If we listen to the PKBSE generated beats we get more of a pulsating marching sound than we do with the beats generated from the static encoding. This is consistent with the groups of subdivisions that increased in amplitude due to the noise from the real device and phase kickback. As a result, we can say that the characteristics of the noise being introduced by real quantum devices are a significant influence on the rhythmic aspects of the system. This could lead to encoding internal rhythms, grooves and feels into the system. This possibly could give quantum computers the ability to feel and understand the concepts of style, groove and personality and creativity in computer/algorithmic music generation.\hfill

\section*{Initial Steps to A Complete Quantum Application}
\medskip
So far we have compared and contrasted the distributions of the database circuits to the results of the timbre register of the input audio track classically. If we increase the amount of qubits in our circuit, or in other words use a larger device we can do this comparison on a quantum computer! The circuit below outlines how this can be done.\hfill

\begin{figure}[htbp]
\begin{center}\vspace{0.1cm}
\includegraphics[scale=0.75]{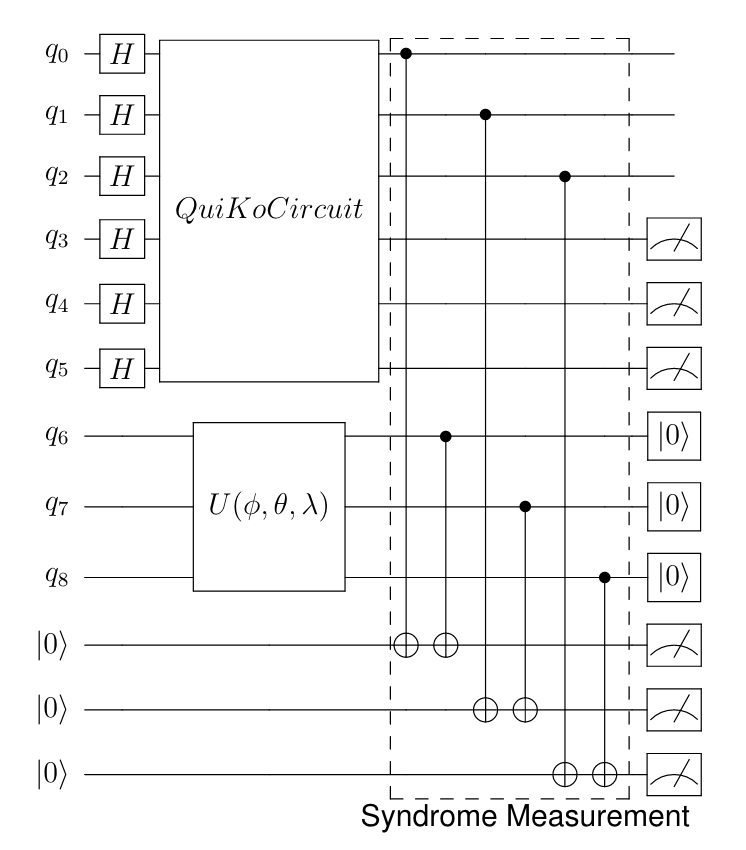}
\caption{Quantum Circuit for comparing the output of the Input audio track to the Audio Tracks in the Database}
\end{center}
\end{figure}

\medskip
Figure 19 shows the case of comparing one track from the audio database to the output of the input audio track for a specific subdivision. To recap, we put the qubits $q_0$ through $q_5$ in superposition by applying $H^{\otimes{n}}$ hadamard gates. We then apply the QuiKo circuit (Static or PKBSE encoding methods) to qubits $q_0$ through $q_5$. We then apply the circuit set up for one track shown in section 4.2. After this a syndrome measurement is implemented to act as a comparator between the QuiKo circuit and each audio database circuit. This will flag a match between the output of the timbre register and the collapsed value of the database circuit. We then see that the qubits on the syndrome measurement are then measured and recorded in a classical register. The spinal cord register on the QuiKo circuit are also measured to record which subdivision the match is associated with.\hfill

\begin{figure}[htbp]
\begin{align*}
\Qcircuit @C=0.50em @R=0.75em{
&\lstick{q_0} & \gate{H} & \multigate{5}{QuiKo Circuit} & \multigate{14}{C_0} & \qw& \multigate{14}{C_3}& \qw & \qw & \multigate{14}{C_1} & \qw & \multigate{14}{C_4}& \qw\\
&\lstick{q_1} & \gate{H} & \ghost{QuiKo Circuit} & \ghost{C_0}& \qw & \ghost{C_3}& \qw & \qw & \ghost{C_1} & \qw & \ghost{C_4}& \qw\\
&\lstick{q_2} & \gate{H} & \ghost{QuiKo Circuit} & \ghost{C_0}& \qw & \ghost{C_3}& \qw & \qw & \ghost{C_1} & \qw & \ghost{C_4}& \qw\\
&\lstick{q_3} & \gate{H} & \ghost{QuiKo Circuit} & \ghost{C_0}& \qw & \ghost{C_3}& \qw & \qw & \ghost{C_1} & \qw & \ghost{C_4}& \meter\\
&\lstick{q_4} & \gate{H} & \ghost{QuiKo Circuit} & \ghost{C_0}& \qw & \ghost{C_3}& \qw & \qw & \ghost{C_1} & \qw & \ghost{C_4}& \meter\\
&\lstick{q_5} & \gate{H} & \ghost{QuiKo Circuit} & \ghost{C_0}& \qw & \ghost{C_3}& \qw & \qw & \ghost{C_1} & \qw & \ghost{C_4}& \meter\\
&\lstick{q_6} & \qw & \multigate{2}{U(\phi,\theta,\lambda)_0} & \ghost{C_0}& \qw & \ghost{C_3}& \gate{\ket{0}} & \multigate{2}{U(\phi,\theta,\lambda)_1} & \ghost{C_1} & \qw & \ghost{C_4}& \qw\\
&\lstick{q_7} & \qw & \ghost{U(\phi,\theta,\lambda)_0} & \ghost{C_0} & \qw& \ghost{C_3}& \gate{\ket{0}} & \ghost{U(\phi,\theta,\lambda)_1} & \ghost{C_1} & \qw & \ghost{C_4}& \qw\\
&\lstick{q_8} & \qw & \ghost{U(\phi,\theta,\lambda)_0} & \ghost{C_0}& \qw & \ghost{C_3}& \gate{\ket{0}} & \ghost{U(\phi,\theta,\lambda)_1} & \ghost{C_1} & \qw & \ghost{C_4}& \qw\\
&\lstick{q_9} & \qw & \multigate{2}{U(\phi,\theta,\lambda)_3} & \ghost{C_0}& \qw & \ghost{C_3}& \gate{\ket{0}} & \multigate{2}{U(\phi,\theta,\lambda)_4} & \ghost{C_1} & \qw & \ghost{C_4}& \qw\\
&\lstick{q_{11}} & \qw & \ghost{U(\phi,\theta,\lambda)_3} & \ghost{C_0} & \qw& \ghost{C_3}& \gate{\ket{0}} & \ghost{U(\phi,\theta,\lambda)_4} & \ghost{C_1} & \qw & \ghost{C_4}& \qw\\
&\lstick{q_{12}} & \qw & \ghost{U(\phi,\theta,\lambda)_3} & \ghost{C_0} & \qw& \ghost{C_3}& \gate{\ket{0}} & \ghost{U(\phi,\theta,\lambda)_4} & \ghost{C_1} & \qw & \ghost{C_4}& \qw\\
&\lstick{q_{13}} & \qw & \qw & \ghost{C_0} & \meter & \ghost{C_3}& \meter & \qw & \ghost{C_1} & \meter & \ghost{C_4} & \meter\\
&\lstick{q_{14}}  & \qw & \qw & \ghost{C_0} & \meter &\ghost{C_3}& \meter & \qw & \ghost{C_1} & \meter & \ghost{C_4} & \meter\\
&\lstick{q_{15}}  & \qw & \qw & \ghost{C_0} & \meter & \ghost{C_3}& \meter & \qw & \ghost{C_1} & \meter & \ghost{C_4} & \meter\\
}
\end{align*}
\caption{Quantum Circuit for Comparing four Audio Tracks to the Input Track}
\end{figure}
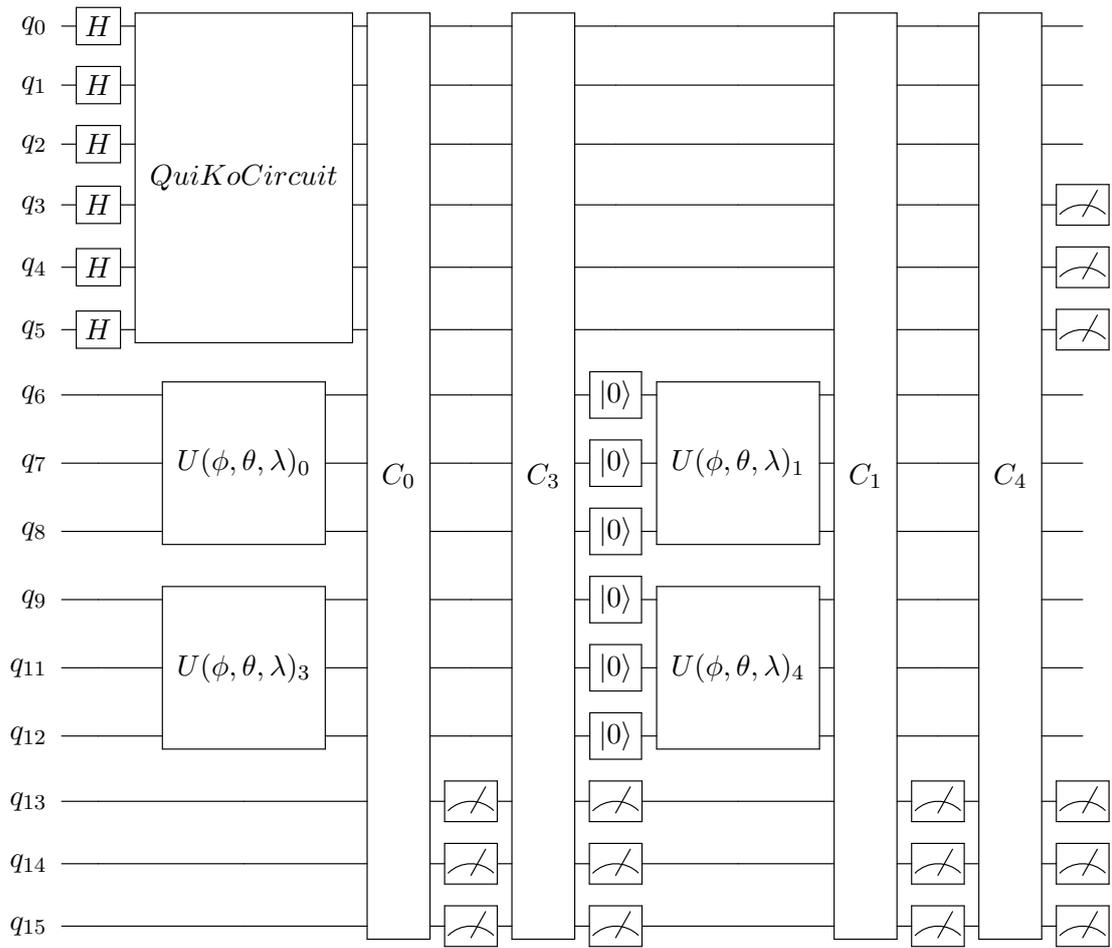

\medskip
Once the matches on the syndrome qubit register is measured then we store the results in a larger classical bit register in their assigned position within the classical bit string. In this case of figure above, the first three bit will be allocated to the syndrome measure between the QuiKo timbre register and the state of the first audio track quantum register. The next three bits will be for the the next syndrome measure with the QuiKo circuit and the second audio track in the database, and so on and so forth. The last three bits of the classical register will be allocated for the subdivision the comparison is happening on, so when we do the post processing or parsing of the results we know how to associate which comparison distribution goes with which subdivision in the measure.\hfill

\medskip
The syndrome measurement is implemented here as and alternative to the more expensive comparators used in various applications for quantum audio signal processing [10][11]. Here compare register is initialized in $\ket{0}$ and then use a CNOT gate to entangle $q_0$ and $q_9$. If $q_0$ happens to be '1' then $q_9$ will flip to '1', and if $q_6$ happens to match then it will flip it back to 0. This also occurs if both $q_0$ and $q_9$ are matching '0's since the CNOT will not trigger. As a result, if we follow this logic for all the qubits in the comparator if we get the compare register to be '000' then the input and the audio track have a match for that particular shot, and since we measure the compare register right after we can reuse it to compare another audio track. We also have to reset the audio track qubits after measurement of the comparator if we want to reuse it for another audio track in the database. Figure next illustrates an example of a full circuit implementation of comparing from the database.\hfill

\begin{figure}[htbp]
\begin{center}\vspace{0.1cm}
\includegraphics[scale=0.75]{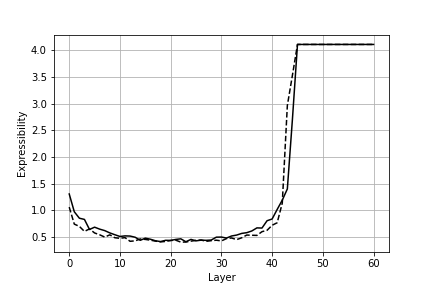}
\caption{Expressibility results for the full circuit including database selection}
\end{center}
\end{figure}
\FloatBarrier
\medskip
If we compare the expressibility metric from the one obtained classically we see that it generally shares the same shape. However, we do see for both the static and PKBSE methods that it hits a maximum around layer 45 and then maintains constant expressibility value of 4.0, which tells us that there is only one outcome for a variety input parameters. In other word, the system is no longer flexible between layers 45 and 61. This is due to the decoherence and quantum volume of the actual device (IBMQ Brooklyn). This becomes a factor due to the fact that we are implementing the circuit in figure 20 for 61 audio tracks. This makes our full circuit very large and the amount of gates and time it takes to process probably approaches or exceeds the quantum volume of IBMQ Brooklyn. In addition, since the qubits in timbre register are not being reset, the qubits within the register decohere over time, which explains why we see a constant flat top after the $45^{th}$ layer in figure 21.\hfill

\section*{Future Work}
\medskip
This experiment has only taken the initial steps in using quantum computers in creating responsive beats. Keep in mind here we only conducted this experiment with one kind of database containing a limited number of samples and variability. In future studies this experiment should be repeated with databases of different samples, lengths and instruments/timbres to truly get a better picture of how these algorithms are performing.\hfill

\medskip
The experiments performed in this chapter only dealt with initially setting both qubit registers in equal superposition. Further investigation is required to know how the system would perform if the initial states of the qubits are not equal. These initial states will be referred to as internal pulse of the system. Different functions and probability distributions can be used as internal pulse states, thus allowing for real world musical rhythms and grooves (i.e. Afro-Cuban Rhythms, Funk, Swing, etc.) to be encoded into the system. Figure * illustrates the change from initializing the timbre and spinal cord registers from superposition to different states.\hfill

\begin{figure}[htbp]
\begin{center}\vspace{0.1cm}
\includegraphics[scale=0.5]{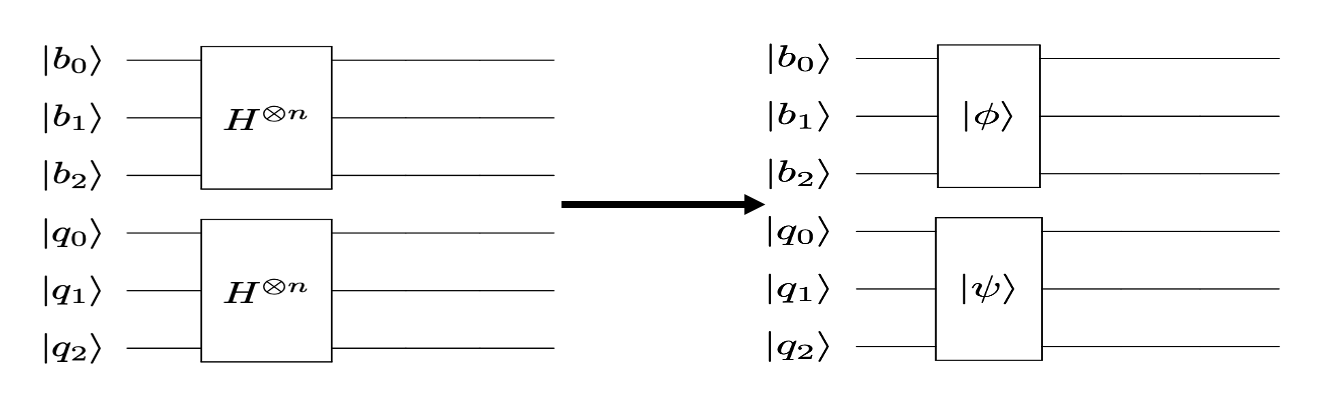}
\caption{Internal Bounce States}
\end{center}
\end{figure}
\FloatBarrier

\medskip
In addition, the ineffectiveness of this methodology is that we still classically compare and contrast the different quantum states from the database. Further research will investigate how to design this process to work within one quantum circuit, and will also look into applying Quantum Access Memory (QRAM) [17] to better store the quantum states of the audio files. It will also be extended to generate segments larger than one measure at a time and to study how to take advantage of elements such as de-coherence between for musical purposes.\hfill

\medskip
Finally, further studies will need to be conducted to increase the resolution of the system. So far we have only dealt with eight note subdivisions. The number of qubits will have be scaled up for account for anything shorter than an eighth note. Initial experiments have been attempted to run these algorithms for more qubits allowing for more sub-bands and subdivisions to be considered. However, as the size of the qubit registers scaled up so did the run time and it became very inefficient. As a result, the method presented in this chapter will have to be adapted for these larger high definition circuits.\hfill

\newpage


\end{document}